\documentclass[11pt, a4paper]{article}

\usepackage{jheppub, slashed, color}
\usepackage{multirow}
\usepackage{tikz}
\usepackage{natbib}
\usepackage{amsmath}
\usepackage{color}
\usepackage[multiple]{footmisc}
\usepackage{textcomp}
\usepackage{amssymb}
\usepackage{mathrsfs}
\usetikzlibrary{positioning}

\DeclareFontFamily{U}{MnSymbolC}{}
\DeclareSymbolFont{MnSyC}{U}{MnSymbolC}{m}{n}
\DeclareFontShape{U}{MnSymbolC}{m}{n}{
    <-6>  MnSymbolC5
   <6-7>  MnSymbolC6
   <7-8>  MnSymbolC7
   <8-9>  MnSymbolC8
   <9-10> MnSymbolC9
  <10-12> MnSymbolC10
  <12->   MnSymbolC12}{}
\DeclareMathSymbol{\intprod}{\mathbin}{MnSyC}{'270}

\newcommand{\Tr}{\mathop{\mathrm{Tr}}\nolimits}

\newcommand{\Z}{\mathbb{Z}}

\newcommand{\R}{\mathbb{R}}
\newcommand{\C}{\mathbb{C}}

\let\nc\newcommand
\let\renc\renewcommand
\nc{\wbar}{\overline}
\let\td\tilde
\let\wtd\widetilde
\let\wht\widehat
\let\mcl\mathcal

\nc{\ab}{{\bar{a}}} \nc{\at}{\tilde{a}} \nc{\ah}{\hat{a}}
\nc{\bb}{{\bar{b}}} \nc{\bt}{\tilde{b}} \nc{\bh}{\hat{b}}
\nc{\cb}{{\bar{c}}} \nc{\ct}{\tilde{c}} 
\nc{\db}{{\bar{d}}} \nc{\dt}{\tilde{d}} \renc{\dh}{\hat{d}}
\nc{\eb}{{\bar{e}}} \nc{\et}{\tilde{e}} \nc{\eh}{\hat{e}}
\nc{\fb}{{\bar{f}}} \nc{\ft}{\tilde{f}} \nc{\fh}{\hat{f}}
\nc{\gb}{{\bar{g}}} \nc{\gt}{\tilde{g}} \nc{\gh}{\hat{g}}
\nc{\hb}{{\bar{h}}} \nc{\hh}{\hat{h}} 
\nc{\ib}{{\bar{\imath}}} \nc{\ih}{\hat{\imath}} 
\nc{\jb}{{\bar{\jmath}}} \nc{\jt}{\tilde{\jmath}} \nc{\jh}{\hat{\jmath}}
\nc{\kb}{{\bar{k}}} \nc{\kt}{\tilde{k}} \nc{\kh}{\hat{k}}
\nc{\lb}{{\bar{l}}} \nc{\lt}{\tilde{l}} \nc{\lh}{\hat{l}}
\nc{\mb}{{\bar{m}}} \nc{\mt}{\tilde{m}} \nc{\mh}{\hat{m}}
\nc{\nb}{{\bar{n}}} \nc{\nt}{\tilde{n}} \nc{\nh}{\hat{n}}
\nc{\ob}{{\bar{o}}} \nc{\ot}{\tilde{o}} \nc{\oh}{\hat{o}}
\nc{\pb}{{\bar{p}}} \nc{\pt}{\tilde{p}} \nc{\ph}{\hat{p}}
\nc{\qb}{{\bar{q}}} \nc{\qt}{\tilde{q}} \nc{\qh}{\hat{q}}
\nc{\rb}{{\bar{r}}} \nc{\rt}{\tilde{r}} \nc{\rh}{\hat{r}}
\renc{\sb}{{\bar{s}}} \nc{\st}{\tilde{s}} \nc{\sh}{\hat{s}}
\nc{\tb}{{\bar{t}}} \renc{\th}{\hat{t}} 
\nc{\ub}{{\bar{u}}} \nc{\ut}{\tilde{u}} \nc{\uh}{\hat{u}}
\nc{\vb}{{\bar{v}}} \nc{\vt}{\tilde{v}} \nc{\vh}{\hat{v}}
\nc{\wb}{{\bar{w}}} \nc{\wt}{\tilde{w}} \nc{\wh}{\hat{w}}
\nc{\xb}{{\bar{x}}} \nc{\xt}{\tilde{x}} \nc{\xh}{\hat{x}}
\nc{\yb}{{\bar{y}}} \nc{\yt}{\tilde{y}} \nc{\yh}{\hat{y}}
\nc{\zb}{{\bar{z}}} \nc{\zt}{\tilde{z}} \nc{\zh}{\hat{z}}

\nc{\Ab}{\wbar{A}} \nc{\At}{\wtd{A}} \nc{\Ah}{\wht{A}}
\nc{\Bb}{\wbar{B}} \nc{\Bt}{\wtd{B}} \nc{\Bh}{\wht{B}}
\nc{\Cb}{\wbar{C}} \nc{\Ct}{\wtd{C}} \nc{\Ch}{\wht{C}}
\nc{\Db}{\wbar{D}} \nc{\Dt}{\wtd{D}} \nc{\Dh}{\wht{D}}
\nc{\Eb}{\wbar{E}} \nc{\Et}{\wtd{E}} \nc{\Eh}{\wht{E}}
\nc{\Fb}{\wbar{F}} \nc{\Ft}{\wtd{F}} \nc{\Fh}{\wht{F}}
\nc{\Gb}{\wbar{G}} \nc{\Gt}{\wtd{G}} \nc{\Gh}{\wht{G}}
\nc{\Hb}{\wbar{H}} \nc{\Ht}{\wtd{H}} \nc{\Hh}{\wht{H}}
\nc{\Ib}{\wbar{I}} \nc{\It}{\wtd{I}} \nc{\Ih}{\wht{I}}
\nc{\Jb}{\wbar{J}} \nc{\Jt}{\wtd{J}} \nc{\Jh}{\wht{J}}
\nc{\Kb}{\wbar{K}} \nc{\Kt}{\wtd{K}} \nc{\Kh}{\wht{K}}
\nc{\Lb}{\wbar{L}} \nc{\Lt}{\wtd{L}} \nc{\Lh}{\wht{L}}
\nc{\Mb}{\wbar{M}} \nc{\Mt}{\wtd{M}} \nc{\Mh}{\wht{M}}
\nc{\Nb}{\wbar{N}} \nc{\Nt}{\wtd{N}} \nc{\Nh}{\wht{N}}
\nc{\Ob}{\wbar{O}} \nc{\Ot}{\wtd{O}} \nc{\Oh}{\wht{O}}
\nc{\Pb}{\wbar{P}} \nc{\Pt}{\wtd{P}} \nc{\Ph}{\wht{P}}
\nc{\Qb}{\wbar{Q}} \nc{\Qt}{\wtd{Q}} \nc{\Qh}{\wht{Q}}
\nc{\Rb}{\wbar{R}} \nc{\Rt}{\wtd{R}} \nc{\Rh}{\wht{R}}
\nc{\Sb}{\wbar{S}} \nc{\St}{\wtd{S}} \nc{\Sh}{\wht{S}}
\nc{\Tb}{\wbar{T}} \nc{\Tt}{\wtd{T}} \nc{\Th}{\wht{T}}
\nc{\Ub}{\wbar{U}} \nc{\Ut}{\wtd{U}} \nc{\Uh}{\wht{U}}
\nc{\Vb}{\wbar{V}} \nc{\Vt}{\wtd{V}} \nc{\Vh}{\wht{V}}
\nc{\Wb}{\wbar{W}} \nc{\Wt}{\wtd{W}} \nc{\Wh}{\wht{W}}
\nc{\Xb}{\wbar{X}} \nc{\Xt}{\wtd{X}} \nc{\Xh}{\wht{X}}
\nc{\Yb}{\wbar{Y}} \nc{\Yt}{\wtd{Y}} \nc{\Yh}{\wht{Y}}
\nc{\Zb}{\wbar{Z}} \nc{\Zt}{\wtd{Z}} \nc{\Zh}{\wht{Z}}

\nc{\CA}{\mcl{A}} \nc{\CAb}{\wbar{\CA}} \nc{\CAt}{\wtd{\CA}} \nc{\CAh}{\wht{\CA}}
\nc{\CB}{\mcl{B}} \nc{\CBb}{\wbar{\CB}} \nc{\CBt}{\wtd{\CB}} \nc{\CBh}{\wht{\CB}}
\nc{\CC}{\mcl{C}} \nc{\CCb}{\wbar{\CC}} \nc{\CCt}{\wtd{\CC}} \nc{\CCh}{\wht{\CC}}
\nc{\cD}{\mcl{D}} \nc{\cDb}{\wbar{\cD}} \nc{\cDt}{\wtd{\cC}} \nc{\cDh}{\wht{\cD}}
\nc{\CE}{\mcl{E}} \nc{\CEb}{\wbar{\CE}} \nc{\CEt}{\wtd{\CE}} \nc{\CEh}{\wht{\CE}}
\nc{\CF}{\mcl{F}} \nc{\CFb}{\wbar{\CF}} \nc{\CFt}{\wtd{\CF}} \nc{\CFh}{\wht{\CF}}
\nc{\CG}{\mcl{G}} \nc{\CGb}{\wbar{\CG}} \nc{\CGt}{\wtd{\CG}} \nc{\CGh}{\wht{\CG}}
\nc{\CH}{\mcl{H}} \nc{\CHb}{\wbar{\CH}} \nc{\CHt}{\wtd{\CH}} \nc{\CHh}{\wht{\CH}}
\nc{\CI}{\mcl{I}} \nc{\CIb}{\wbar{\CI}} \nc{\CIt}{\wtd{\CI}} \nc{\CIh}{\wht{\CI}}
\nc{\CJ}{\mcl{J}} \nc{\CJb}{\wbar{\CJ}} \nc{\CJt}{\wtd{\CJ}} \nc{\CJh}{\wht{\CJ}}
\nc{\CK}{\mcl{K}} \nc{\CKb}{\wbar{\CK}} \nc{\CKt}{\wtd{\CK}} \nc{\CKh}{\wht{\CK}}
\nc{\CL}{\mcl{L}} \nc{\CLb}{\wbar{\CL}} \nc{\CLt}{\wtd{\CL}} \nc{\CLh}{\wht{\CL}}
\nc{\CM}{\mcl{M}} \nc{\CMb}{\wbar{\CM}} \nc{\CMt}{\wtd{\CM}} \nc{\CMh}{\wht{\CM}}
\nc{\CN}{\mcl{N}} \nc{\CNb}{\wbar{\CN}} \nc{\CNt}{\wtd{\CN}} \nc{\CNh}{\wht{\CN}}
\nc{\CO}{\mcl{O}} \nc{\COb}{\wbar{\CO}} \nc{\COt}{\wtd{\CO}} \nc{\COh}{\wht{\CO}}
\nc{\CQ}{\mcl{Q}} \nc{\CQb}{\wbar{\CQ}} \nc{\CQt}{\wtd{\CQ}} \nc{\CQh}{\wht{\CQ}}
\nc{\CR}{\mcl{R}} \nc{\CRb}{\wbar{\CR}} \nc{\CRt}{\wtd{\CR}} \nc{\CRh}{\wht{\CR}}
\nc{\CS}{\mcl{S}} \nc{\CSb}{\wbar{\CS}} \nc{\CSt}{\wtd{\CS}} \nc{\CSh}{\wht{\CS}}
\nc{\CT}{\mcl{T}} \nc{\CTb}{\wbar{\CT}} \nc{\CTt}{\wtd{\CT}} \nc{\CTh}{\wht{\CT}}
\nc{\CU}{\mcl{U}} \nc{\CUb}{\wbar{\CU}} \nc{\CUt}{\wtd{\CU}} \nc{\CUh}{\wht{\CU}}
\nc{\CV}{\mcl{V}} \nc{\CVb}{\wbar{\CV}} \nc{\CVt}{\wtd{\CV}} \nc{\CVh}{\wht{\CV}}
\nc{\CW}{\mcl{W}} \nc{\CWb}{\wbar{\CW}} \nc{\CWt}{\wtd{\CW}} \nc{\CWh}{\wht{\CW}}
\nc{\CX}{\mcl{X}} \nc{\CXb}{\wbar{\CX}} \nc{\CXt}{\wtd{\CX}} \nc{\CXh}{\wht{\CX}}
\nc{\CY}{\mcl{Y}} \nc{\CYb}{\wbar{\CY}} \nc{\CYt}{\wtd{\CY}} \nc{\CYh}{\wht{\CY}}
\nc{\CZ}{\mcl{Z}} \nc{\CZb}{\wbar{\CZ}} \nc{\CZt}{\wtd{\CZ}} \nc{\CZh}{\wht{\CZ}}

\let\eps\epsilon
\let\ups\upsilon
\let\veps\varepsilon
\let\vtht\vartheta
\let\vsgm\varsigma
\let\vphi\varphi
\let\vrho\varrho

\nc{\alphab}{\bar{\alpha}} \nc{\alphat}{\td{\alpha}} \nc{\alphah}{\hat{\alpha}}
\nc{\betab}{\bar{\beta}}   \nc{\betat}{\td{\beta}}   \nc{\betah}{\hat{\beta}} 
\nc{\gammab}{\bar{\gamma}} \nc{\gammat}{\td{\gamma}} \nc{\gammah}{\hat{\gamma}} 
\nc{\deltab}{\bar{\delta}} \nc{\deltat}{\td{\delta}} \nc{\deltah}{\hat{\delta}} 
\nc{\epsilonb}{\bar{\eps}} \nc{\epsilont}{\td{\eps}} \nc{\epsilonh}{\hat{\eps}} 
\nc{\vepsb}{\bar{\veps}}   \nc{\vepst}{\td{\veps}}   \nc{\vepsh}{\hat{\veps}} 
\nc{\zetab}{\bar{\zeta}}   \nc{\zetat}{\td{\zeta}}   \nc{\zetah}{\hat{\zeta}} 
\nc{\etab}{\bar{\eta}}     \nc{\etat}{\td{\eta}}     \nc{\etah}{\hat{\eta}} 
\nc{\thetab}{\bar{\theta}} \nc{\thetat}{\td{\theta}} \nc{\thetah}{\hat{\theta}} 
\nc{\vthetab}{\bar{\vtht}} \nc{\vthetat}{\td{\vtht}} \nc{\vthetah}{\hat{\vtht}} 
\nc{\lambdab}{\bar{\lambda}} \nc{\lambdat}{\td{\lambda}} \nc{\lambdah}{\hat{\lambda}} 
\nc{\iotab}{\bar{\iota}}   \nc{\iotat}{\td{\iota}}   \nc{\iotah}{\hat{\iota}} 
\nc{\kappab}{\bar{\kappa}} \nc{\kappat}{\td{\kappa}} \nc{\kappah}{\hat{\kappa}} 
\nc{\lmdb}{\bar{\lmd}}     \nc{\lmdt}{\td{\lmd}}     \nc{\lmdh}{\hat{\lmd}} 
\nc{\mub}{\bar{\mu}}       \nc{\mut}{\td{\mu}}       \nc{\muh}{\hat{\mu}} 
\nc{\nub}{\bar{\nu}}       \nc{\nut}{\td{\nu}}       \nc{\nuh}{\hat{\nu}} 
\nc{\xib}{\bar{\xi}}       \nc{\xit}{\td{\xi}}       \nc{\xih}{\hat{\xi}} 
\nc{\pib}{\bar{\pi}}       \nc{\pit}{\td{\pi}}       \nc{\pih}{\hat{\pi}} 
\nc{\vpib}{\bar{\vpi}}     \nc{\vpit}{\td{\vpi}}     \nc{\vpih}{\hat{\vpi}} 
\nc{\rhob}{\bar{\rho}}     \nc{\rhot}{\td{\rho}}     \nc{\rhoh}{\hat{\rho}} 
\nc{\vrhob}{\bar{\vrho}}   \nc{\vrhot}{\td{\vrho}}   \nc{\vrhoh}{\hat{\vrho}} 
\nc{\sigmab}{\bar{\sigma}} \nc{\sigmat}{\td{\sigma}} \nc{\sigmah}{\hat{\sigma}} 
\nc{\vsigmab}{\bar{\vsgm}} \nc{\vsigmat}{\td{\vsgm}} \nc{\vsigmah}{\hat{\vsgm}} 
\nc{\taub}{\bar{\tau}}     \nc{\taut}{\td{\tau}}     \nc{\tauh}{\hat{\tau}} 
\nc{\upsb}{\bar{\ups}} \nc{\upst}{\td{\ups}} \nc{\upsh}{\hat{\ups}} 
\nc{\phib}{\bar{\phi}}     \nc{\phit}{\td{\phi}}     \nc{\phih}{\hat{\phi}} 
\nc{\varphib}{\bar{\vphi}}   \nc{\varphit}{\td{\vphi}}   \nc{\varphih}{\hat{\vphi}} 
\nc{\chib}{\bar{\chi}}     \nc{\chit}{\td{\chi}}     \nc{\chih}{\hat{\chi}} 
\nc{\psib}{\bar{\psi}}     \nc{\psit}{\wtd{\psi}}     \nc{\psih}{\hat{\psi}} 
\nc{\omegab}{\bar{\omega}} \nc{\omegat}{\td{\omega}} \nc{\omegah}{\hat{\omega}} 

\nc{\Gammab}{\wbar{\Gamma}}     \nc{\Gammat}{\wtd{\Gamma}}     \nc{\Gammah}{\wht{\Gamma}}
\nc{\Deltab}{\wbar{\Delta}}     \nc{\Deltat}{\wtd{\Delta}}     \nc{\Deltah}{\wht{\Delta}}
\nc{\Thetab}{\wbar{\Theta}}     \nc{\Thetat}{\wtd{\Theta}}     \nc{\Thetah}{\wht{\Theta}}
\nc{\Lambdab}{\wbar{\Lambda}}   \nc{\Lambdat}{\wtd{\Lambda}}   \nc{\Lambdah}{\wht{\Lambda}}
\nc{\Xib}{\wbar{\Xi}}           \nc{\Xit}{\wtd{\Xi}}           \nc{\Xih}{\wht{\Xi}}
\nc{\Pib}{\wbar{\Pi}}           \nc{\Pit}{\wtd{\Pi}}           \nc{\Pih}{\wht{\Pi}}
\nc{\Sigmab}{\wbar{\Sigma}}     \nc{\Sigmat}{\wtd{\Sigma}}     \nc{\Sigmah}{\wht{\Sigma}}
\nc{\Upsilonb}{\wbar{\Upsilon}} \nc{\Upsilont}{\wtd{\Upsilon}} \nc{\Upsilonh}{\wht{\Upsilon}}
\nc{\Phib}{\wbar{\Phi}}         \nc{\Phit}{\wtd{\Phi}}         \nc{\Phih}{\wht{\Phi}}
\nc{\Psib}{\wbar{\Psi}}         \nc{\Psit}{\wtd{\Psi}}         \nc{\Psih}{\wht{\Psi}}
\nc{\Omegab}{\wbar{\Omega}}     \nc{\Omegat}{\wtd{\Omega}}     \nc{\Omegah}{\wht{\Omega}}

\title{
 Little Strings, Quasi-topological Sigma Model on Loop Group,  and Toroidal Lie Algebras  }
\author[]{Meer Ashwinkumar,}
\author[]{Jingnan Cao,}
\author[]{Yuan Luo,}
\author[]{Meng-Chwan Tan}
\author[]{and Qin Zhao}

\emailAdd{meerashwinkumar@u.nus.edu}
\emailAdd{jingnan.cao@u.nus.edu}
\emailAdd{phyluoy@nus.edu.sg}
\emailAdd{mctan@nus.edu.sg}
\emailAdd{phyzhq@nus.edu.sg}
\affiliation[]{Department of Physics, National University of
  Singapore \\  2 Science Drive 3, Singapore 117551}
\abstract{ We study the 
ground states and left-excited states of the $A_{k-1}$ $\mathcal{N}= (2,0)$ little string theory.
Via a theorem by Atiyah \cite{Atiyah1984}, 
 these sectors can be captured by a supersymmetric nonlinear sigma model on $\C P^1$ with target space the based loop group of $ SU(k)$. The 
 ground states, described by $L^2$-cohomology classes, form modules over an affine Lie algebra, while the 
 left-excited states, described by chiral differential operators, form modules over a toroidal Lie algebra. We also apply our results to analyze the 1/2 and 1/4 BPS sectors of the M5-brane worldvolume theory.
}
\dedicated{In loving memory of See-Hong Tan.}


\arxivnumber{}

\renewcommand{\ell}{l}

\begin{document}

\maketitle
\flushbottom

\section{Introduction}

 Little string theories (LST) are non-gravitational string theories that exist in six spacetime dimensions which reduce to interacting local quantum field theories in the 
limit where the string length scale $l_s\rightarrow 0$. In particular, the $A_{k-1}$ $\mathcal{N}=(2,0)$ little string theory flows in the $l_s\rightarrow 0$ limit to the $A_{k-1}$ $\mathcal{N}=(2,0)$ superconformal field theory (SCFT). This SCFT does not have a classical Lagrangian description because it contains a non-abelian two-form potential with a self-dual three-form field strength, whence it has no known (conformal) Lagrangian formulation {\cite{Witten1997}}. This SCFT has recently provided us with several dualities of quantum field theories as well as connections to mathematics, implying that its corresponding LST ought to be at least just as rich. 

The $A_{k-1}$ $\mathcal{N}=(2,0)$ LST can be understood to arise as follows. 
Consider a stack of $k$ space-filling NS5-branes in type IIA string theory on $\R^{9,1}$.  The NS5-branes break half of the 32 spacetime supersymmetries such that $\mathcal{N}=(2,0)$ supersymmetry is preserved on the worldvolume.  Lightcone coordinates can be used for the worldvolume of the stack, and
one of the lightlike coordinates 
can be compactified to a circle of radius $R$, i.e., the NS5-brane worldvolume topology is  $S^1_- \times \R_+ \times \R^4$. 
Here, $S^1_- \times \R_+$ can be understood as $(S^1\times \R)^{1,1}$ (where $S^1$ is a small spacelike circle and $\R$ is timelike), with $S^1$ boosted by a large amount \cite{Seiberg1997}. Now, an F-string can wrap the circle $S^1$ 
with $N$ units of discrete momentum and winding number $m$. If we take the 
limit where the string coupling $g_s\rightarrow 0$ while keeping  $l_s=\sqrt{\alpha'}$ finite, the F-string will be bound to the branes, and we obtain $A_{k-1}$ $\mathcal{N}=(2,0)$ little string theory on $(S^1\times \R)^{1,1}\times \R^4$. 

The $A_{k-1}$ $\mathcal{N}=(2,0)$ little string theory can be described as follows. Now, T-duality takes us to type IIB string theory, whereby $N$ is the winding number and $m$ is the momentum number of the F-string around a lightlike circle of radius $\alpha'/R$ (or equivalently, a very large spacelike circle). Subsequently, performing S-duality leads us to a 
bound state of $k$ D5-branes and a D-string which winds the  circle $N$ times. 
In the low energy limit, 
the D5-brane worldvolume theory is an $SU(k)$ gauge theory,\footnote{We have conveniently frozen the center-of-mass dynamics of the stack of D5-branes, reducing the gauge symmetry of the low energy worldvolume theory from $U(k)$ to $SU(k)$.
} and the D-string appears as an $SU(k)$ $N$-instanton in the $\R^4$ subspace of 
the worldvolume. Then, in the very low energy 
discrete lightcone quantization (DLCQ) limit,
the theory on the D-string worldsheet is that of the $\mathcal{N}=(4,4)$ supersymmetric sigma model with target space being the moduli space of $SU(k)$ $N$-instantons on $\R^4$ (denoted as $\mathcal{M}_{SU(k)}^N(\R^4)$) {\cite{Aharony1997,Aharony1886,Aharony1999}}. In other words, the DLCQ of the $A_{k-1}$ $\mathcal{N}=(2,0)$ little string theory with $N$ units of discrete momentum can be understood in terms of the aforementioned sigma model. 

 In this paper, we investigate the topological and quasi-topological sectors of this $\mathcal{N} = (4,4)$ supersymmetric sigma model in an attempt to understand the ground states and left-excited states of the little string theory, respectively. Our strategy would be to investigate the aforementioned sectors via an auxiliary sigma model on $\C P^1$ which has $\mathcal{N}$=(2,2) supersymmetry and target space $\Omega SU(k)$, the based loop group of $SU(k)$.\footnote{The reason this auxiliary sigma model has $\mathcal{N}$=(2,2) supersymmetry instead of $\mathcal{N}$=(4,4) supersymmetry is because the $\Omega SU(k)$ target space is (as we shall explain in the next section) a K{\"a}hler manifold, whereas the moduli space of $SU(k)$ instantons on $\R^4$ is a hyperk{\"a}hler manifold.} The use of this auxiliary sigma model is possible because of a mathematical theorem of Atiyah's {\cite{Atiyah1984}}, which states that the moduli space $\mathcal{M}_{G}^N(\R^4)$ of $N$-instantons on $\R^4$ for gauge group $G$ is diffeomorphic to the moduli space $\mathcal{M}(\C P^1 \xrightarrow[hol.]{N}\Omega{G})$ of $N$-degree holomorphic maps from $\C P^1$ to the based loop group, $\Omega G$. 
A summary and plan of the paper is as follows. In Section 2, we briefly review some mathematical facts about loop groups which we shall use, and describe Atiyah's theorem in detail. In Section 3, we shall introduce the supersymmetric A-twisted nonlinear sigma model on $\C P^1$ with $\Omega SU(k)$ target space and explain its topological and quasi-topological sectors, elucidating its properties. We shall demonstrate the appearance of current algebras in both sectors, namely the toroidal Lie algebra $\mathfrak{su}(k)_{\textrm{tor}}$ and the affine Lie algebra $\mathfrak{su}(k)_{\textrm{aff}}$ in the quasi-topological and topological sectors, 
respectively. In Section 4, we explain the correspondence between the topological and quasi-topological sectors of the loop group sigma model and that of the instanton moduli space sigma model. In turn, this will allow us to understand the 
 ground and left-excited sectors of the $A_{k-1}$ $\mathcal{N}=(2,0)$ little string theory, 
and show that the local observables in both these sectors form modules over their respective current algebras. We then arrive at a sigma model derivation of Braverman and Finkelberg's conjecture for the case of $SU(k)$ instantons on $\R^4$, i.e., that the intersection cohomology of the moduli space of $SU(k)$ instantons on $\R^4$ forms modules over the affine Lie algebra $\mathfrak{su}(k)_{\textrm{aff}}$, and a generalization thereof, i.e., that the \v{C}ech cohomology of the sheaf of certain chiral differential operators on the moduli space of $SU(k)$ instantons on $\R^4$ forms modules over the toroidal Lie algebra  $\mathfrak{su}(k)_{\textrm{tor}}$.
In Section 5, we explain the correspondence between the topological and quasi-topological sectors of the  $\Omega SU(k)$ sigma model and the 1/2 BPS and 1/4 BPS sectors of the M5-brane worldvolume theory. Then, we calculate the partition functions of both the topological and quasi-topological sectors, and thereby the partition functions of the 1/2 BPS and 1/4 BPS sectors of the M5-brane worldvolume theory.\footnote{Note that the 1/4 BPS sector of the M5-brane worldvolume theory is different from the 1/4 BPS sector of its low energy $A_{k-1}$ $\mathcal{N}=(2,0)$ SCFT. The DLCQ of the $A_{k-1}$ $\mathcal{N}=(2,0)$ SCFT with $N$ units of momentum can be described by the $\mathcal{N}=(4,4)$ supersymmetric quantum mechanics on $\mathcal{M}_{SU(k)}^N(\R^4)$ \cite{Aharony1997}, 
whereby the local observables in the 1/4 BPS sector -- which are invariant under the 4 anti-chiral supercharges of the $\mathcal{N}=(4,4)$ supersymmetric quantum mechanics -- are holomorphic forms on $\mathcal{M}_{SU(k)}^N(\R^4)$.  As any holomorphic form is automatically harmonic with respect to the Kahler metric on a Kahler manifold,  they correspond to $L^2$-harmonic forms on $\mathcal{M}_{SU(k)}^N(\R^4)$. On the other hand, the 1/4 BPS sector of the M5-brane worldvolume theory is described by chiral differential operators.}
\section{Loop Groups ($LG$) and Atiyah's Theorem}
A loop group {\cite{pressley1986loop}} is the group consisting of maps from the unit circle $S^1$ to a group $G$:
\begin{equation}
f:S^1\rightarrow G,
\end{equation}
and is denoted as $LG$. We can parametrize the unit circle via $t=e^{i\theta}$. The group composition law is inherited from the composition law of the group $G$, taken pointwise for every value of $\theta$. If we assume that $G$ is a 
Lie group, an element of $LG$ (connected to its identity) can be denoted as
\begin{equation}\label{expmap}
e^{i\lambda_a(t)T^a}=e^{i\lambda_{am}T^ae^{im\theta}},
\end{equation}
where $T^a$ is an element of the Lie algebra $\mathfrak{g}$ corresponding to $G$, and $\lambda_a(t)$ is a real-valued function of $S^1$.

The loop algebra $L\mathfrak{g}$ consists of maps from $S^1$ to the Lie algebra $\mathfrak{g}$. Each element of the loop algebra is an element of $\mathfrak{g}$, as well as a Laurent polynomial\footnote{A Laurent polynomial is a linear combination of both positive and negative powers of the variable $t$, with finitely many nonzero coefficients, all valued in $\C$. The set of Laurent polynomials is closed under multiplication and addition, and hence forms the Laurent polynomial ring, $\C[t,t^{-1}]$.} in the variable $t=e^{i\theta}$, i.e.,
\begin{equation}
L\mathfrak{g}=\mathfrak{g} \otimes  \C[t,t^{-1}],
\end{equation}
 and thus $\lambda\in L\mathfrak{g}$ can be written as $\lambda=\lambda_a(t)T^a=\lambda_{am}T^ae^{im\theta}$, where $m\in\Z$. Elements of the loop algebra satisfy the Lie bracket
\begin{equation}
[T^a\otimes e^{im\theta},T^b\otimes e^{in\theta}]=if^{ab}_c T^c\otimes e^{i(m+n)\theta},
\end{equation} 
or equivalently
\begin{equation}\label{loopalgebraT}
[T^{am},T^{bn}]=if^{ab}_c T^{c(m+n)},
\end{equation} 
where $T^{am}\equiv T^a\otimes e^{im\theta}$.

One can also define the based loop group $\Omega G$, if one imposes the based point condition 
\begin{equation}\label{basepoint}
f(\theta=0)=I
\end{equation}
on the maps, where $I$ denotes the identity element of $G$. This is 
a subgroup of $LG$.
One may notice that $G$ is a subgroup of $LG$ as well,
consisting of constant maps. Since \eqref{basepoint} only allows the identity as  the image of a constant map in $\Omega G$, we have
\begin{equation}
\Omega G \cong LG/G,
\end{equation}
i.e., it is a homogeneous space.
In fact, $\Omega G$ is an infinite-dimensional manifold. Let $\xi$ and $\eta$ be elements of $\Omega\mathfrak{g}$, the based loop algebra. Then, expanding them in the $L\mathfrak{g}$ basis gives
\begin{equation}\label{expandd}
\begin{aligned}
\xi(\theta)&= \xi_{n} e^{in\theta}= \xi_{an} T^a e^{in\theta},\\
\eta(\theta)&=\eta_n e^{in\theta}=\eta_{an} T^a e^{in\theta}.
\end{aligned}
\end{equation}
The based point condition $\eqref{basepoint}$, which can be written as $e^{i\xi(\theta=0)}=1$, then translates to $\sum_n \xi_{an} T^a=0$ at the Lie algebra level.

$\Omega G$ admits another description as a homogeneous space, 
\begin{equation}
\Omega G \cong LG_{\mathbb{C}}/L^+G_{\mathbb{C}},
\end{equation}
where $LG_{\mathbb{C}}$ denotes the group consisting of maps from $S^1$ to the complexification of $G$, denoted $G_{\mathbb{C}}$, whilst $L^+G_{\mathbb{C}}$ is the space of boundary values of holomorphic maps from the unit disk in $\mathbb{C}$ to $G_{\mathbb{C}}$. It is this identification that endows $\Omega G$ with a complex structure. In fact, one may embed $\Omega G$ in $LG_{\mathbb{C}}$. Locally, this can be understood as follows. Let $\kappa \in L\mathfrak{g}_{\mathbb{C}}$, where $\mathfrak{g}_{\mathbb{C}}$ denotes the complexification of $\mathfrak{g}$. Imposing the based point condition 
\begin{equation}
\sum_n \kappa_{an}T^a=0
\end{equation}
reduces $\kappa$ to an element of $\Omega\mathfrak{g}_{\mathbb{C}}$, and imposing the condition
\begin{equation}\label{phiphibar}
\kappa_{(-n)}=\overline{\kappa}_{n}
\end{equation}
(where $\kappa_{(-n)}=\kappa_{a(-n)}T^a$ and $\overline{\kappa}_{n}=\overline{\kappa}_{an}(T^a)^\dag$) reduces $\kappa$ to an element of $\Omega\mathfrak{g}$ \cite{sergeev2010kahler}. $\Omega G$ also admits a closed nondegenerate two-form $\omega$, i.e., it has a symplectic structure. The complex and symplectic structures of $\Omega G$ are compatible, and conspire to make it an infinite-dimensional K{\"a}hler manifold \cite{Hu1991,pressley1986loop}. 

The complex structure $J$, and symplectic structure $\omega$, can be combined to define the metric of $\Omega G$: 
\begin{equation}
g(\xi,\eta)=\omega(\xi,J\eta).
\end{equation}
In components, this is given {\cite{Hu1991}} by 
\begin{equation}
g_{am,bn}=|n|\delta_{n+m,0}\Tr(T_a T_b).
\end{equation}
The Christoffel symbols and Riemann curvature tensor can be calculated for this metric, and for this we refer the reader to \cite{freed1988,freed1985}.

Atiyah's theorem {\cite{Atiyah1984}} is an insightful theorem linking instantons in 4d and 2d. Its precise statement is that, for any classical group $G=SU(k)$, $Sp(k)$ or $SO(k)$ and positive integer $N$, the parameter space $\mathcal{M}_{G}^N(\R^4)$ (or $\mathcal{M}_{G}^N(S^4)$) of Yang-Mills $N$-instantons over $\mathbb{R}^4$ (or its conformal compactification $S^4$) with gauge group $G$ modulo based gauge transformations is diffeomorphic to the parameter space $\mathcal{M}(\C P^1 \xrightarrow[hol.]{N}\Omega{G})$ of all based holomorphic maps from $\C P^1$ to $\Omega G$ of degree $N$. By based gauge transformations, we mean gauge transformations which tend to $1$ at $\infty\in \mathbb{R}^4$ (or $\infty\in S^4$), and by based holomorphic maps, we mean holomorphic maps which map $\infty\in \mathbb{C}P^1$ to $1 \in \Omega G$ (i.e., $\lim_{z\to\infty} e^{i\xi_{an}(z)T^{an}}=1$). We can write the theorem succinctly as 
\begin{equation}
\mathcal{M}_{G}^N(\R^4) \cong\mathcal{M}(\C P^1 \xrightarrow[hol.]{N}\Omega{G}).
\end{equation}
The pullback of the K{\"a}hler two-form $\omega$ via the based holomorphic maps $\Phi$ defines a differential form on the worldsheet $\mathbb{C}P^1$. Also, for simple $G$, $H_2(\Omega G,\Z)\cong\Z$, and the integral of the pullback 
provides the degree of holomorphic map:
\begin{equation}\label{holmapdegree}
N=\textrm{degree}(\Phi)=\int\limits_{\mathbb{C}P^1} \Phi^* \omega.
\end{equation}

\section{Supersymmetric A-twisted Sigma Model on $\C P^1$ with $\Omega SU(k)$ Target Space}
\subsection{The A-model Action, Supersymmetries and Local Observables}
We begin with an exposition on the two-dimensional supersymmetric nonlinear sigma model with target space $\Omega SU(k)$ and worldsheet $\C P^1$. 
It is convenient to first construct this as as a sigma model governing maps 
\begin{equation}
\Phi : \C P^1 \rightarrow \Omega SU(k)_{\mathbb{C}}
\end{equation}
(where $SU(k)_{\mathbb{C}}$ is the complexification of $SU(k)$), and to then obtain $\Omega SU(k)$ as a subspace of $\Omega SU(k))_{\mathbb{C}}$ via an algebraic constraint.
Picking local coordinates $z, \overline{z}$ on $\C P^1$, and $\phi^{am}, \overline{\phi}^{a\overline{m}}$ on $\Omega SU(k))_{\mathbb{C}}$, the map $\Phi$ can then be described locally by the fields $\phi^{am}(z,\overline{z})$
and $\overline{\phi}^{a\overline{m}}(z,\overline{z})$, where $a=1,\ldots, \textrm{dim }SU(k)$ and $m$, $\overline{m}\in \Z$.
\footnote{The fields $\phi^{am}$ and $\overline{\phi}^{a\overline{m}}$ are analogous to the Lie algebra parameters $\lambda_{am}$ in \eqref{expmap}. 
} The remaining fields of the model are the following smooth sections of fiber bundles on $\C P^1$:
\begin{equation}
\begin{aligned}
\psi_+^{am} &\in \Gamma(K^\frac{1}{2} \otimes \Phi^*T\Omega SU(k)_{\mathbb{C}}),\\
\overline{\psi}_+^{a\overline{m}} &\in \Gamma(K^\frac{1}{2}\otimes\Phi^*\overline{T\Omega SU(k)_{\mathbb{C}}}),\\
\psi_-^{am} &\in \Gamma(\overline{K}^\frac{1}{2}\otimes\Phi^*T\Omega SU(k)_{\mathbb{C}}),\\
\overline{\psi}_-^{a\overline{m}} &\in \Gamma(\overline{K}^\frac{1}{2} \otimes \Phi^*\overline{T\Omega SU(k)_{\mathbb{C}}}),
\end{aligned}
\end{equation}
where $T\Omega SU(k)_{\mathbb{C}}$ and $\overline{T\Omega SU(k)_{\mathbb{C}}}$ are the holomorphic and anti-holomorphic tangent bundles of $\Omega SU(k)_{\mathbb{C}}$, and where $K^\frac{1}{2}$ and $\overline{K}^\frac{1}{2}$ are the positive and negative chirality spinor bundles of $\C P^1$.

Let $g_{am,b\overline{n}}$ be the metric on $\Omega SU(k)_{\mathbb{C}}$. The action is then given by\footnote{We have chosen a flat Hermitian metric $(\eta_{z\overline{z}}=\frac{1}{2})$ on the worldsheet, since every two-dimensional metric is conformally flat, and can be Weyl rescaled to be flat locally (recall that any Lagrangian density is only defined locally).}
\begin{equation}\label{action0}
\begin{aligned}
S&=\int d^2z \Big(  g_{am,b\overline{n}}(\frac{1}{2}\partial_z\phi^{am}\partial_{\overline{z}}\overline{\phi}^{b\overline{n}}+\frac{1}{2}\partial_{\overline{z}}\phi^{am}\partial_{z}\overline{\phi}^{b\overline{n}}+\overline{\psi}_-^{b\overline{n}}D_{\overline{z}}\psi_-^{am}+\psi_+^{am}D_{z}\overline{\psi_+}^{b\overline{n}})
\\&-R_{am,c\overline{p},bn,d\overline{q}}\psi^{am}_+\psi_-^{bn}\overline{\psi}^{c\overline{p}}_{-}\overline{\psi}_+^{d\overline{q}}\Big),
\end{aligned}
\end{equation}
where $m, n,\overline{n},\overline{p},\overline{q}\in \Z$, 
and where the covariant derivatives are 
\begin{equation}
  \begin{aligned}
D_{\overline{z}}\psi_-^{am}&=\partial_{\overline{z}}\psi_-^{am}  + \Gamma^{am}_{bn,cp}\partial_{\overline{z}}\phi^{bn}\psi_-^{cp},
\\
D_{z}\overline{\psi}_+^{a\overline{m}}&=\partial_{z}\overline{\psi}_+^{a\overline{m}}  + \Gamma^{a\overline{m}}_{b\overline{n},c\overline{p}}\partial_{z}\overline{\phi}^{b\overline{n}}\overline{\psi}_+^{c\overline{p}}.
  \end{aligned}
\end{equation}

Both barred and unbarred indices in the action \eqref{action0} sum over the set of integers, and it shall be convenient for our purposes to replace the barred indices by unbarred indices, e.g., $\overline{\phi}^{b\overline{n}}\rightarrow\overline{\phi}^{bn}$. To avoid ambiguity, the connection on $\overline{T\Omega SU(k)_{\mathbb{C}}}$ is renotated as
\begin{equation}
\Gamma^{a\overline{m}}_{b\overline{n},c\overline{p}}\rightarrow\overline{\Gamma}^{am}_{bn,cp}.
\end{equation}
%
The reduction of $\Omega SU(k)_{\mathbb{C}}$ to $\Omega SU(k)$ is achieved by imposing the constraint \eqref{phiphibar} on the coordinates and tangent vectors of $\Omega SU(k)_{\mathbb{C}}$
\begin{equation}
\begin{aligned}
\phi^{a(-n)}&=\overline{\phi}^{an},\\
\psi_{\mp}^{a(-n)}&=\overline{\psi}_{\pm}^{an},
\end{aligned},
\end{equation}
where we have taken into account the hermiticity of the generators of $SU(k)$. These constraints precisely  reduce the number of degrees of freedom by half, and shall always be assumed hereafter.

We are after the local observables of the topological and quasi-topological sectors of this sigma model, both of which are accessible via a 
`twist'. 
This is done by redefining the generator $M_E$ of $U(1)$ Euclidean rotations to be $M'_E=M_E+R$, where $R$ is a generator of a $U(1)$ R-symmetry of the action \eqref{action0}. There are two ways to do this, using either the $U(1)$ vector R-symmetry or $U(1)$ axial R-symmetry, and this leads to the A-model and B-model \cite{Witten1991}. We shall use the A-twist which leads to the A-model, since, as we shall see in later sections, this choice will eventually enable us to make contact with the physics of the 
$A_{k-1}$ $\mathcal{N}=(2,0)$ little string theory . 

The twisting does not affect the spins of the bosonic fields, but the fermionic fields become the following smooth sections of fiber bundles on $\C P^1$:
\begin{equation}
\begin{aligned}
\psi_+^{am}\rightarrow\rho_{\overline{z}}^{am} &\in \Gamma(\overline{K} \otimes \Phi^*T\Omega SU(k)),\\
\overline{\psi}_+^{am}\rightarrow\overline{\chi}^{am} &\in \Gamma(\Phi^*\overline{T\Omega SU(k)}),\\
\psi_-^{am}\rightarrow\chi^{am} &\in \Gamma(\Phi^*T\Omega SU(k)),\\
\overline{\psi}_-^{am}\rightarrow\overline{\rho}_z^{am} &\in \Gamma(K \otimes \Phi^*\overline{T\Omega SU(k)}),
\end{aligned}
\end{equation}
where $K$ and $\overline{K}$ are the canonical and anti-canonical bundles of $\C P^1$ (i.e., bundles of one-forms of types (1,0) and (0,1)). The A-model action is 
\begin{equation}\label{action}
\begin{aligned}
S&=\int d^2z \Big(  g_{am,bn}(\frac{1}{2}\partial_z\phi^{am}\partial_{\overline{z}}\overline{\phi}^{bn}+\frac{1}{2}\partial_{\overline{z}}\phi^{am}\partial_{z}\overline{\phi}^{bn}+\overline{\rho}^{bn}_{z}D_{\overline{z}}\chi^{am}+\rho^{am}_{\overline{z}}D_{z}\overline{\chi}^{bn})
\\&-R_{cp,bn,dq,am}\overline{\rho}^{cp}_{z}\chi^{bn}\overline{\chi}^{dq}\rho^{am}_{\overline{z}} \Big)\\
&=\int d^2z \Big(  g_{am,bn}(\partial_{\overline{z}}\phi^{am}\partial_{z}\overline{\phi}^{bn}+\overline{\rho}^{bn}_{z}D_{\overline{z}}\chi^{am}+\rho^{am}_{\overline{z}}D_{z}\overline{\chi}^{bn})
\\&-R_{cp,bn,dq,am}\overline{\rho}^{cp}_{z}\chi^{bn}\overline{\chi}^{dq}\rho^{am}_{\overline{z}}+\int \Phi^*\omega\\&
=S_{pert.}+\int \Phi^*\omega,
\end{aligned}
\end{equation}
where
\begin{equation}
  \begin{aligned}
D_{\overline{z}}\chi^{am}&=\partial_{\overline{z}}\chi^{am}  + \Gamma^{am}_{bn,cp}\partial_{\overline{z}}\phi^{bn}\chi^{cp},
\\
D_{z}\overline{\chi}^{am}&=\partial_{z}\overline{\chi}^{am}  + \overline{\Gamma}^{am}_{bn,cp}\partial_{z}\overline{\phi}^{bn}\overline{\chi}^{cp},
  \end{aligned}
\end{equation}
$S_{pert.}$ denotes the pertubative action, and $\Phi^*\omega$ is the pullback of the K{\"a}hler form of $\Omega SU(k)$.
The A-model action $\eqref{action}$ is invariant under the following supersymmetries, generated by the scalar supercharges $\overline{Q}_+$ and $Q_-$:  
\begin{equation}\label{trans}
  \begin{aligned}
    \delta \phi^{am} &= \epsilon_+\chi^{am},
    \\
    \delta \overline{\phi}^{am} &= \overline{\epsilon}_-\overline{\chi}^{am},
    \\
    \delta \rho^{am}_{\overline{z}} &=-\overline{\epsilon}_-\partial_{\overline{z}}\phi^{am}-\epsilon_+\Gamma^{am}_{bn,cp}\chi^{bn}\rho^{cp}_{\overline{z}},
    \\
 \delta \overline{\rho}^{am}_{z} &=- \epsilon_+\partial_{z}\overline{\phi}^{am}-\overline{\epsilon}_-\overline{\Gamma}^{am}_{bn,cp}\overline{\chi}^{bn}\overline{\rho}_{z}^{cp},
    \\
  \delta\chi^{am}
    &= 0,
   \\
   \delta \overline{\chi}^{am}
    &= 0,      
  \end{aligned}
\end{equation}
where $\delta=\overline{\epsilon}_-\overline{Q}_+ + \epsilon_+Q_-$, and $\delta^2=0$ is satisfied on-shell.

The rest of this section is devoted to understanding the fully-twisted and half-twisted versions of our A-model, with emphasis on mathematical descriptions of the local quantum observables of both models, which shall be useful for us in the following sections.
The fully-twisted A-model \cite{Witten1988,Witten1991} is conformal at the classical level since the energy-momentum tensor is traceless ($T_{z\overline{z}}=0$), leading to it having holomorphic ($T_{zz}=T(z)$) and antiholomorphic ($T_{\overline{z}\overline{z}}=\overline{T}(\overline{z})$) nonzero components. Both these nonzero components $T_{zz}$ and $T_{\overline{z}\overline{z}}$ are $Q_A=(\overline{Q}_+ +Q_-)$-exact, which means that the A-model is in fact topological. The half-twisted A-model {\cite{witten2007two,Tan2006}} corresponds to the sector wherein one only considers the supercharge $\overline{Q}_+$, and the supersymmetry transformations are those of $\eqref{trans}$ with $\epsilon_+=0$. It is still conformal at the classical level, since the energy-momentum tensor remains the same. The crucial difference is that now $T_{\overline{z}\overline{z}}$ is $\overline{Q}_+$-exact, but $T_{zz}$ is not, and hence the model is not topological. We refer to this model as the quasi-topological model.



When computing correlation functions of 
observables in the topological $\Omega SU(k)$ sigma model, one may use the fact that the periods of the K{\"a}hler form $\omega$ are integers (c.f. \eqref{holmapdegree}),
\begin{equation}\label{instaper}
\int_{\C P^1}\Phi^*\omega=N,
\end{equation}
to express correlation functions as \cite{Witten1991}
\begin{equation}\label{corrfunc0}
\langle\prod_{\gamma} \mathcal{O}_{\gamma}\rangle=\sum_N e^{-N}\int_{F_N}\mathcal{D}\phi\mathcal{D}\overline{\phi}\mathcal{D}\rho_{\overline{z}}\mathcal{D}\overline{\rho}_z\mathcal{D}\chi\mathcal{D}\overline{\chi}e^{-S_{pert.}} \prod_{\gamma}\mathcal{O}_{\gamma},
\end{equation}
where an explicit factor of $e^{-N}$ has been pulled out in each term on the right.\footnote{To be precise, one also needs to include auxiliary fields in the action, in order to obtain $\delta^2=0$ off-shell. Details on this are presented in equation \eqref{equivalentaction} and below.} Here $F_N$ denotes the component of field space corresponding to maps of degree $N$, and the components of the path integral measure are defined as $\mathcal{D}X=(\mathcal{D}X^{ak}\mathcal{D}X^{bl}\mathcal{D}X^{cn}\ldots)$. The observables $\mathcal{O}_{\gamma}$ are understood to be those which are in the $Q_A$-cohomology, since supersymmetry at the quantum level requires $\langle\{Q_A,O\}\rangle=0$ for any observable $O$.

Since $R_{cp,bn,dq,am}=g_{cp,ek}R^{ek}_{\textrm{   }bn,dq,am}$, the target space metric is an overall factor in the Lagrangian density of the action \eqref{action}, and hence gives rise to an infinite number of coupling constants, via its Taylor expansion. It can be shown that, pertubatively, the topological model remains invariant when rescaling these couplings \cite{Witten1991}. The argument is as follows. The action \eqref{action} can be written as\footnote{The expression \eqref{locaction} only holds modulo terms that vanish using the $\rho$ equations of motion, but it can be made to hold off-shell by modifying the supersymmetry transformations of $\rho$ \cite{Witten1991}.  }
\begin{equation}\label{locaction}
S=\int d^2z \{Q_A,V'\}+\int\Phi^*\omega,
\end{equation}
where
\begin{equation}
V'=g_{am,bn}\Big( \overline{\rho}^{bn}_{z}\partial_{\overline{z}}\phi^{am}+\partial_z\overline{\phi}^{bn}\rho^{am}_{\overline{z}}\Big)
\end{equation}
Multiplying $g_{am,bn}$ by a factor $t$, \eqref{corrfunc0} becomes
\begin{equation}\label{corrfuncttop}
\langle\prod_{\gamma} \mathcal{O}_{\gamma}\rangle=\sum_N e^{-tN}\int_{F_N}\mathcal{D}\phi\mathcal{D}\overline{\phi}\mathcal{D}\rho_{\overline{z}}\mathcal{D}\overline{\rho}_z\mathcal{D}\chi\mathcal{D}\overline{\chi}e^{-t\int d^2z \{Q_A,V'\}} \prod_{\gamma}\mathcal{O}_{\gamma},
\end{equation}
where
\begin{equation}
\frac{d}{dt}\Big(\int_{F_N}\mathcal{D}\phi\mathcal{D}\overline{\phi}\mathcal{D}\rho_{\overline{z}}\mathcal{D}\overline{\rho}_z\mathcal{D}\chi\mathcal{D}\overline{\chi}e^{-t\int d^2z \{Q_A,V'\}} \prod_{\gamma}\mathcal{O}_{\gamma}   \Big)=\langle \{Q_A,\ldots\}\rangle=0,
\end{equation}
i.e., the path integral over $F_N$ is independent of the value that $t$ takes.

In particular, for each path integral in \eqref{corrfuncttop}, one may take the weak-coupling or infinite-volume limit where $t\rightarrow \infty$, whereupon 
the contributions to the path integral localize to fluctuations around the following classical saddle point configuration which minimizes the first term of \eqref{locaction}:
\begin{equation}\label{BPS}
\begin{aligned}
\partial_{\overline{z}}\phi^{am}&=0.
\end{aligned}
\end{equation}
We shall refer to this as the BPS condition, and its solutions are holomorphic maps from $\C P^1$ to $\Omega SU(k)$ (also known as worldsheet instantons). The degree of holomorphic map (or worldsheet instanton number) is given by \eqref{instaper}, and terms in \eqref{corrfuncttop} corresponding to $N<0$ vanish, as there are no holomorphic maps of negative degree.
The quantum fluctuations of the fields around these classical solutions, represented by the fermionic and bosonic one-loop determinants, cancel exactly due to supersymmetry. As a result, path integrals in the topological A-model reduce to a sum over worldsheet instanton sectors (labelled by $N$) of ordinary, finite-dimensional integrals over the moduli space $\mathcal{M}(\C P^1 \xrightarrow[hol.]{N} \Omega SU(k))$ of holomorphic maps of degree $N$ from $\C P^1$ to $\Omega SU(k)$, with 
an overall factor of $e^{-tN}$ multiplying each integral. Since the local observables one considers in this model are those which belong to the $Q_A$-cohomology,
the $Q_A$-closure and non-$Q_A$-exactness of these observables implies one-to-one correspondence with de Rham cohomology classes on $\mathcal{M}(\C P^1 \xrightarrow[hol.]{N} \Omega SU(k))$;\footnote{The identification of $Q_A$ with the coboundary operator $d$ on $\Omega SU(k)$ is a natural consequence of \eqref{trans} (with $\epsilon_+$=$\overline{\epsilon}_-$=1), since the supersymmetry transformations of the coordinates $\phi^{am}$, $\overline{\phi}^{a\overline{m}}$ of $\Omega SU(k)$ give us the fields $\chi^{am}$, $\overline{\chi}^{a\overline{m}}$ which are Grassmannian, whose anticommuting products can be identified with wedge products of one-forms on $\Omega SU(k)$. 
As a result, the supersymmetry transformations of worldsheet $N$-instantons (which satisfy \eqref{BPS}) give us Grassmannian fields which transform as one-forms on $\mathcal{M}(\C P^1 \xrightarrow[hol.]{N} \Omega SU(k))$ \cite{Witten1988}, allowing $Q_A$ to be identified with $d$ on $\mathcal{M}(\C P^1 \xrightarrow[hol.]{N} \Omega SU(k))$.
Moreover, $Q_A^2=0$ always holds for sigma models on closed worldsheets, unless one has a pure $\mathcal{N}=(0,2)$ sigma model \cite{tan2008chiral}. } however, since $\mathcal{M}(\C P^1 \xrightarrow[hol.]{N} \Omega SU(k))$ is noncompact, we should identify the local observables with $L^2$-harmonic forms, which give rise to $L^2$-cohomology.

Also, using the fact that $T(z)$ and $\overline{T}(\overline{z})$ are $Q_A$-exact, one can show that local observables with nonzero holomorphic and antiholomorphic conformal dimensions are trivial in $Q_A$-cohomology. The argument is as follows. A local operator $\mathcal{O}$ inserted at the origin has conformal dimension $(n,m)$ if under the rescaling $z\rightarrow\lambda z$, $\overline{z}\rightarrow\overline{\lambda}\overline{z}$ (which is a symmetry of our theory since $T_{z\overline{z}}=0$), it transforms as $\partial^{n+m}/\partial z^n\partial z^m$, i.e., as 
\begin{equation}\label{confdime}
\mathcal{O}\rightarrow\lambda^{-n}\overline{\lambda}^{-m}\mathcal{O}, 
\end{equation}
where $n$ and $m$ are positive integers. However, only local operators with $m=n=0$ survive in $Q_A$-cohomology. The reason for the previous statement is that the rescalings of $z$ and $\overline{z}$ are generated by $L_0=\oint dz z T_{zz}$ and $\overline{L}_0=\oint d\bar{z} \bar{z} T_{\overline{z}\overline{z}}$ respectively. As noted previously, $T_{zz}$ and $T_{\overline{z}\overline{z}}$ are $Q_A$-exact, so $L_0+\overline{L}_0=\{Q_A,V_0\}$ for some $V_0$. If $\mathcal{O}$ is to be allowed as a local physical operator, it must at least be true that $\{Q_A,\mathcal{O}\}=0$. Subsequently, we have $[(L_0+\overline{L}_0),\mathcal{O}]=\{Q_A,[V_0,\mathcal{O}]\}$. On account of the eigenvalue of $L_0$ and $\overline{L}_0$ on $\mathcal{O}$ being $m$ and $n$ respectively, we have $[(L_0+\overline{L}_0),\mathcal{O}] =(m+n)\mathcal{O}$. Therefore, if $m\neq 0$ or $n\neq 0$, it is true that $\mathcal{O}$ is $Q_A$-exact and as such trivial in $Q_A$-cohomology. Consequently, the local observables of the topological A-model can only have holomorphic and antiholomorphic conformal dimensions equal to zero, and hence correspond to ground states, via the state-operator isomorphism.

The action in the quasi-topological model can be cast into the form
\begin{equation}\label{locaction2}
S=\int d^2z \{\overline{Q}_+,W'\} + \ldots +\int \Phi^*\omega ,
\end{equation}
where $W'$ is a metric-dependent combination of fields, and where the ellipsis indicates additional terms which are metric-independent, but depend on the complex structure of the target space.\footnote{This shall be expounded on further in Section 3.2.} Since the metric-dependence of the pertubative action is contained entirely in a $\overline{Q}_+$-exact term, the 
path integral over a particular component of field space, $F_N$, is independent of the couplings generated by the metric of the target space, i.e., multiplying $g_{am,bn}$ by $t$, a correlation function of observables in the $\overline{Q}_+$-cohomology has the form
\begin{equation}
\langle\prod_{\gamma} \widetilde{\mathcal{O}}_{\gamma}\rangle=\sum_N e^{-tN}\int_{F_N}\mathcal{D}\phi\mathcal{D}\overline{\phi}\mathcal{D}\rho_{\overline{z}}\mathcal{D}\overline{\rho}_z\mathcal{D}\chi\mathcal{D}\overline{\chi}e^{-\int d^2z (\{\overline{Q}_+,W'(t)\}+\ldots)} \prod_{\gamma}\widetilde{\mathcal{O}}_{\gamma},
\end{equation}
where
\begin{equation}\label{tindep}
\frac{d}{dt}\Big(\int_{F_N}\mathcal{D}\phi\mathcal{D}\overline{\phi}\mathcal{D}\rho_{\overline{z}}\mathcal{D}\overline{\rho}_z\mathcal{D}\chi\mathcal{D}\overline{\chi}e^{-\int d^2z (\{\overline{Q}_+,W'(t)\}+\ldots)} \prod_{\gamma}\widetilde{\mathcal{O}}_{\gamma}   \Big)=\langle \{\overline{Q}_+,\ldots\}\rangle=0,
\end{equation}
which means that the path integral over $F_N$ is independent of the value that $t$ takes.
 
Similar to the topological case, one may take the weak-coupling or infinite-volume limit where $t\rightarrow \infty$; contributions to the path integral then localize to fluctuations around the saddle point configuration $\eqref{BPS}$
, and the path integral reduces to a sum over worldsheet instanton sectors (labelled by $N$) of finite-dimensional integrals over $\mathcal{M}(\C P^1 \xrightarrow[hol.]{N} \Omega SU(k))$, upon cancellation of the fermionic and bosonic one-loop determinants. The third term of \eqref{locaction2} contributes an overall factor of $e^{-tN}$ to each integral, and terms corresponding to $N<0$ vanish, as in the topological case. As one might expect, the local observables one considers belong to the $\overline{Q}_+$-cohomology. However, since $T_{\overline{z}\overline{z}}$ is $\overline{Q}_+$-exact but not $T_{zz}$, it can be shown that these observables must have zero antiholomorphic conformal dimension, but may have nonzero holomorphic conformal dimension (unlike the topological model), using arguments analogous to those of the previous paragraph \cite{Tan2006}. Hence, we find that the quasi-topological model not only contains ground states, but also contains left-excited states, via the state-operator isomorphism. Therefore, the excited states of the quasi-topological model correspond to half of the excited states of the untwisted sigma model on $\C P^1$ with $\Omega SU(k)$ target space.

Furnishing a purely mathematical description of the local observables of the quasi-topological model is not as straightforward as in the topological case.
It is known that the half-twisted A-model can be described purely (without using the mathematically nonrigorous path integral) via the theory of chiral differential operators (CDO's) {\cite{kapustin2005chiral,witten2007two,Tan2006,frenkel2007mirror}}. In particular, for this half-twisted version of the $\mathcal{N}=(2,2)$ supersymmetric sigma model, the corresponding CDO's are the chiral de Rham complex \cite{kapustin2005chiral,Tan2006}. 
As explained before, after half-twisting the A-model, only one scalar supercharge $\overline{Q}_+$ remains. Now, even at the classical level, the $\overline{Q}_+$-cohomology 
cannot be described purely by the ordinary Dolbeault or $\bar{\partial}$-cohomology, and perturbative corrections only serve to strengthen this deviation. It is found that \v{C}ech cohomology can be used to describe the perturbative sheaf of $\overline{Q}_+$-cohomology {\cite{witten2007two,Tan2006}}. The local operators and local observables all belong to \v{C}ech cohomology. Moreover
, in an anomaly-free sigma model, a globally defined sheaf of chiral algebras can also be constructed without obstruction in \v{C}ech cohomology.
\footnote{The sheaf of chiral algebras is 
defined locally, 
and refers to the chiral algebra, OPE and the chiral ring in every open set of the manifold. } The above statements are all in perturbative expansion; however, in our sigma model from $\C P^1$ to $\Omega SU(k)$, our main focus will be on higher degree maps rather than the degree zero maps. 
Therefore, we go to an auxiliary $\mathcal{N}=(2,2)$ sigma model whose target space is $\mathcal{M}(\C P^1 \xrightarrow[hol.]{N} \Omega SU(k))$, and consider its pertubative sector, 
whereby the correlation functions of our theory and the correlation functions of the auxiliary theory are identical, with the former multiplied by the constant $e^{-N}$. Thus, we can still identify the physical observables  of our quasi-topological model with well-defined mathematical CDO's. 
In particular, the local observables of the quasi-topological model are described by the \v{C}ech cohomology of the sheaf of chiral de Rham complex on $\mathcal{M}(\C P^1 \xrightarrow[hol.]{N} \Omega SU(k))$.

In Section 4, we shall use the descriptions of local observables of both the topological and quasi-topological models in terms of cohomology classes defined on $\mathcal{M}(\C P^1 \xrightarrow[hol.]{N} \Omega SU(k))$ to describe the 
ground and left-excited sectors of the 
$A_{k-1}$ $\mathcal{N}=(2,0)$ little string theory, via Atiyah's theorem.
However, Atiyah's theorem works for \textit{based} holomorphic maps, and to this end, according to the last paragraph of Section 2, we shall impose an extra constraint on the bosonic scalar field $\phi^{an}$, i.e., it should satisfy $\lim_{z \rightarrow \infty} \phi_{an}(z,\overline{z})T^{an}=0$, which means that $\lim_{z \rightarrow \infty} \phi_{an}(z,\overline{z})=0$, since the $T^{an}$ are linearly independent. 
{\subsection{Global $LSU(k)$ Symmetry and Current Algebra}

Now, let us proceed to study the symmetry of our action \eqref{action}. Since $\Omega SU(k)$ can be understood as the homogeneous space $L SU(k)/SU(k)$, this implies that $\Omega SU(k)$ admits a transitive $L SU(k)$ isometry. This geometrical symmetry of the target space will manifest as a global symmetry of the supersymmetric action \eqref{action}. To grasp how this arises, let us first review how isometries under compact Lie groups manifest in sigma models on finite dimensional K{\"a}hler manifolds. 
\\
\mbox{}\par\nobreak
\noindent
\textit{Isometries of the Target Space in Sigma Models} 

For any supersymmetric nonlinear sigma model, an isometry of its target space, $X$, forms a global symmetry of the action \cite{2012arXiv1207.1241L}. Call the isometry group $G$. An isometry is generated by a set of Killing vector fields, $V^a$, where  $a=1,\ldots,\textrm{dim }G$. The Lie derivative of the metric with respect to $\zeta_aV^a$ (where $\zeta_a$ are a set of real, infinitesimal parameters) vanishes; this is the Killing equation. If $X$ is a K{\"a}hler manifold, the two basic structures it has is a (torsion-free) Hermitian metric and a complex structure, and an isometry should preserve both. In other words, the Lie derivative of the complex structure should also vanish, and this results in Killing vector fields having either holomorphic or antiholomorphic components, i.e., 
\begin{equation}
V^a=\sum_i^n V^{a,i}\frac{\partial}{\partial\phi^i}+\sum_{\overline{\imath}}^n \overline{V}^{a,\overline{\imath}}\frac{\partial}{ \partial\phi^{\overline{\imath}}}
\end{equation}
 ($n=\textrm{dim}_{\mathbb{C}} X$), 
where
\begin{equation}
\frac{\partial V^{a,i}}{\partial \phi^{\overline{\jmath}}}=\frac{\partial \overline{V}^{a,\overline{\imath}}}{\partial \phi^{j}}=0.
\end{equation}
The coordinates ($\phi$) and tangent vectors ($\psi$) of $X$ transform under the isometry as
\begin{equation}\label{gtransform}
\begin{aligned}
\delta \phi^i &= \zeta_a V^{a,i},\\
\delta \overline{\phi}^{\overline{\imath}} &= \zeta_a \overline{V}^{a,\overline{\imath}},\\
\delta \psi^i &= \zeta_a \partial_k (V^{a,i})\psi^k,\\
\delta \overline{\psi}^{\overline{\imath}} &= \zeta_a \partial_{\overline{k}}(\overline{V}^{a,\overline{\imath}})\psi^{\overline{k}}.
\end{aligned}
\end{equation}

The statement that the Killing vector fields generate an action on $X$ implies that each vector field $V^a$ corresponds to an element $T^a$ of the Lie algebra of $G$, and that they realise an antihomomorphism of that Lie algebra, i.e.,
\begin{equation}
[V^a,V^b]=-if^{ab}_c V^c.
\end{equation} 
Locally, this is written explicitly in components as 
\begin{equation}
[V^a,V^b]^i = V^{a,j}(\frac{\partial V^{b,i}}{\partial\phi^j})-V^{b,j}(\frac{\partial V^{a,i}}{\partial\phi^j})=-if^{ab}_c V^{c,i},
\end{equation}
and
\begin{equation}
[\overline{V}^a,\overline{V}^b]^{\overline{\imath}}=\overline{V}^{a,\overline{\jmath}}(\frac{\partial \overline{V}^{b,\overline{\imath}}}{\partial\overline{\phi}^{\overline{\jmath}}})-\overline{V}^{b,\overline{\jmath}}(\frac{\partial \overline{V}^{a,\overline{\imath}}}{\partial\overline{\phi}^{\overline{\jmath}}}) =-if^{ab}_c \overline{V}^{c,\overline{\imath}}.
\end{equation}
Given the A-model action for target space $X$,
\begin{equation}
S_{X}=\int d^2z \Big(  g_{i\overline{\jmath}}(\partial_{\overline{z}}\phi^{i}\partial_{z}\overline{\phi}^{\overline{\jmath}}+\overline{\rho}^{\overline{\jmath}}_{z}D_{\overline{z}}\chi^{i}+\rho^{i}_{\overline{z}}D_{z}\overline{\chi}^{\overline{\jmath}})-R_{\overline{k}j\overline{l}i}\overline{\rho}^{\overline{k}}_{z}\chi^{j}\overline{\chi}^{\overline{l}}\rho^{i}_{\overline{z}}\Big)+\int \Phi^*\omega,
\end{equation}
we know that the bosonic and fermionic fields transform as coordinates and tangent vectors respectively on the target space. Then, varying these fields under the $G$-isometry as in \eqref{gtransform} gives
\begin{equation}\label{Gtransaction}
\begin{aligned}
\delta_G S_{X}&=\int d^2z \Big(  \mathcal{L}_V g_{i\overline{\jmath}}(\partial_{\overline{z}}\phi^{i}\partial_{z}\overline{\phi}^{\overline{\jmath}}+\overline{\rho}^{\overline{\jmath}}_{z}D_{\overline{z}}\chi^{i}+\rho^{i}_{\overline{z}}D_{z}\overline{\chi}^{\overline{\jmath}})+g_{i\overline{\jmath}}\overline{\rho}^{\overline{\jmath}}_{z}\mathcal{L}_V\Gamma^{i}_{jk}\partial_{\overline{z}}\phi^{j}\chi^{k}\\
&+g_{i\overline{\jmath}}\rho^{i}_{\overline{z}}\mathcal{L}_V\Gamma^{\overline{\jmath}}_{\overline{\imath}\overline{k}}\partial_{z}\overline {\phi}^{i}\overline{\chi}^{\overline{k}}-\mathcal{L}_VR_{\overline{k}j\overline{l}i}\overline{\rho}^{\overline{k}}_{z}\chi^{j}\overline{\chi}^{\overline{l}}\rho^{i}_{\overline{z}}\Big)+\int \Phi^*\mathcal{L}_V\omega ,
\end{aligned}
\end{equation}
where $V=\zeta_aV^a$ and $\mathcal{L}_V$ is the Lie derivative with respect to $V$.\footnote{Although the Christoffel symbols are not tensors, and do not have intrinsic geometrical meaning, they have a well defined Lie derivative (see \cite{yano1957theory}, page 8, equation 2.16).} As mentioned above, if $V$ generates an isometry on $X$, then
\begin{equation}\label{liemetric}
\mathcal{L}_Vg_{i\overline{\jmath}}=0.
\end{equation}
Now, note that the Lie derivative of the Christoffel symbol can be expressed solely in terms of the Lie derivative of the metric (\cite{yano1957theory}, page 52, equation 3.1), and the Lie derivative of the Riemann curvature tensor can be expressed solely in terms of the Lie derivative of the Christoffel symbol (\cite{yano1957theory}, page 52, equation 3.2). Additionally, the Lie derivative of the K{\"a}hler 2-form can also be expressed in terms of the Lie derivative of the metric, since the components of the K{\"a}hler form are proportional to the metric. The previous statements, together with \eqref{liemetric}, imply that the transformation of the action \eqref{Gtransaction} under the global symmetry corresponding to the $G$-isometry of the target space is just zero. 
\\
\mbox{}\par\nobreak
\noindent
\textit{The $LSU(k)$ Isometry of the $\Omega SU(k)$ Sigma Model} 

Let us shift our attention back to the $\Omega SU(k)$ sigma model \eqref{action}, and expound on its global $LSU(k)$ symmetry. The Lie algebra for $LSU(k)$ is the loop algebra $L\mathfrak{su}(k)$ \eqref{loopalgebraT}, and each element $T^{am}$ of the loop algebra  corresponds to a Killing vector field $V^{am}$ on  $\Omega SU(k)$ (\cite{freed1988}, page 238). The collection of all these Killing vector fields generate the $LSU(k)$ isometry of $\Omega SU(k)$. Furthermore, there is an antihomorphism from the loop algebra $L\mathfrak{su}(k)$ to these Killing vector fields (\cite{freed1988}, page 240), i.e., they should satisfy 

\begin{equation}\label{loopantihom}
[V^{am},V^{bn}] = -if^{ab}_c V^{c\{m+n\}}.
\end{equation}
In terms of the local coordinate parametrization we have used to describe $\Omega SU(k)$, 
\begin{equation}
V^{bk}=V^{bk,an}\frac{\partial}{\partial \phi^{an}}+\overline{V}^{bk,an}\frac{\partial}{\partial \overline{\phi}^{an}},
\end{equation}
with \eqref{loopantihom} given as
\begin{equation}\label{pde1}
[V^{am},V^{bn}]^{dp} =V^{am,eq}(\frac{\partial V^{bn,dp}}{\partial\phi^{eq}})-V^{bn,eq}(\frac{\partial V^{am,dp}}{\partial\phi^{eq}})= -if^{ab}_c V^{c\{m+n\},dp},
\end{equation}
\begin{equation}\label{pde2}
[\overline{V}^{am},\overline{V}^{bn}]^{dp} =\overline{V}^{am,eq}(\frac{\partial \overline{V}^{bn,dp}}{\partial\overline{\phi}^{eq}})-\overline{V}^{bn,eq}(\frac{\partial \overline{V}^{am,dp}}{\partial\overline{\phi}^{eq}})= -if^{ab}_c \overline{V}^{c\{m+n\},dp}.
\end{equation}
It must also be true that
\begin{equation}\label{holantihol}
\frac{\partial V^{bk,dp}}{\partial \overline{\phi}^{eq}}=\frac{\partial \overline{V}^{bk,dp}}{\partial \phi^{eq}}=0.
\end{equation}

 The coordinates and tangent vectors of $\Omega SU(k)$ transform under the $LSU(k)$ symmetry as 
 \begin{equation}
\begin{aligned}
\delta \phi^{an}&=\sum_{k\in \mathbb{Z}} \zeta^k_{b} V^{bk,an},\\
\delta \overline{\phi}^{an}&=\sum_{k\in \mathbb{Z}} \zeta^k_{b} \overline{V}^{bk,an},\\
\delta \psi^{an} &= \sum_{k\in \mathbb{Z}} \zeta^k_{b} \frac{\partial}{\partial\phi^{dm}} (V^{bk,an})\psi^{dm},\\
\delta \overline{\psi}^{an} &= \sum_{k\in \mathbb{Z}} \zeta^k_{b} \frac{\partial}{\partial\overline{\phi}^{dm}} (\overline{V}^{bk,an})\overline{\psi}^{dm}.
\end{aligned}
\end{equation}
Noting that the bosonic and fermionic fields of the action \eqref{action} transform as coordinates and tangent vectors respectively, the $LSU(k)$ transformation of the action is found to be 
\begin{equation}\label{lgtransformact}
\begin{aligned}
\delta_{LSU(k)}S=&\int d^2z \Big(  \mathcal{L}_Vg_{am,bn}(\partial_{\overline{z}}\phi^{am}\partial_{z}\overline{\phi}^{bn}+\overline{\rho}^{bn}_{z}D_{\overline{z}}\chi^{am}+\rho^{am}_{\overline{z}}D_{z}\overline{\chi}^{bn})\\&+g_{am,bn}\overline{\rho}^{bn}_{z}\mathcal{L}_V\Gamma^{am}_{bn,cp}\partial_{\overline{z}}\phi^{bn}\chi^{cp}+g_{am,bn}\rho^{am}_{\overline{z}}\mathcal{L}_V\overline{\Gamma}^{am}_{bn,cp}\partial_{z}\overline{\phi}^{bn}\overline{\chi}^{cp}
\\&-\mathcal{L}_VR_{cp,bn,dq,am}\overline{\rho}^{cp}_{z}\chi^{bn}\overline{\chi}^{dq}\rho^{am}_{\overline{z}}\Big)+\int \Phi^*\mathcal{L}_V\omega.
\end{aligned}
\end{equation}
The Lie derivative with respect to $V=\sum_k\zeta^k_bV^{bk}$ acting on the $\Omega SU(k)$ metric is zero (\cite{freed1988}, page 240), i.e.,
\begin{equation}\label{lieloopmetric}
\mathcal{L}_Vg_{am,bn}=0.
\end{equation}
This means that the $LSU(k)$ transformation \eqref{lgtransformact} of the action must be zero, using the arguments below equation \eqref{liemetric}. 

 We would now like to find the explicit form of the components of $V^{bk}$  , in order to find the explicit field transformations which leave the action \eqref{action} invariant. The solutions of the partial differential equations
\eqref{pde1} and \eqref{pde2} which satisfy \eqref{holantihol} are given by  
\begin{equation}
\begin{aligned}
V^{bk,an}&=if^{ab}_c \phi^{c\{n-k\}},\\
\overline{V}^{bk,an}&=if^{ab}_c \overline{\phi}^{c\{n-k\}}
\end{aligned}
\end{equation}
The field transformations under the $LSU(k)$ symmetry are therefore given as
\begin{equation}
\begin{aligned}
\delta \phi^{an}&=\sum_{k\in \mathbb{Z}} \zeta^k_{b} V^{bk,an}=\sum_{k\in \mathbb{Z}} if^{ab}_c\zeta^k_{b} \phi^{c\{n-k\}},\\
\delta \overline{\phi}^{an}&=\sum_{k\in \mathbb{Z}} \zeta^k_{b} \overline{V}^{bk,an}=\sum_{k\in \mathbb{Z}} if^{ab}_c\zeta^k_{b} \overline{\phi}^{c\{n-k\}},\\
\delta \rho^{an}_{\overline{z}} &=\sum_{k\in \mathbb{Z}} \zeta^k_{b} \frac{\partial}{\partial\phi^{dm}} (V^{bk,an})\rho_{\overline{z}}^{dm}=\sum_{k\in \mathbb{Z}} if^{ab}_c\zeta^k_{b} \rho_{\overline{z}}^{c\{n-k\}},\\
\delta \overline{\rho}^{an}_z &= \sum_{k\in \mathbb{Z}} \zeta^k_{b} \frac{\partial}{\partial\overline{\phi}^{dm}} (\overline{V}^{bk,an})\overline{\rho}_{z}^{dm}=\sum_{k\in \mathbb{Z}} if^{ab}_c\zeta^k_{b} \overline{\rho}_z^{c\{n-k\}},\\
\delta \chi^{an}&=\sum_{k\in \mathbb{Z}} \zeta^k_{b} \frac{\partial}{\partial\phi^{dm}} (V^{bk,an})\chi^{dm}=\sum_{k\in \mathbb{Z}} if^{ab}_c\zeta^k_{b} \chi^{c\{n-k\}},\\
\delta \overline{\chi}^{an}&=\sum_{k\in \mathbb{Z}} \zeta^k_{b} \frac{\partial}{\partial\overline{\phi}^{dm}} (\overline{V}^{bk,an})\overline{\chi}^{dm}=\sum_{k\in \mathbb{Z}} if^{ab}_c\zeta^k_{b} \overline{\chi}^{c\{n-k\}}.\\
\end{aligned}
\end{equation}
It is beneficial to note that all the fields transform in the same manner, that is, as
\begin{equation}\label{variation}
\delta X^{an}=\sum_{k\in \mathbb{Z}} if^{ab}_c\zeta^k_{b} X^{c\{n-k\}}.
\end{equation}
Having understood the global $LSU(k)$ symmetry of the action \eqref{action}, we shall proceed to show that this classical symmetry is responsible for the appearance of a current algebra in the quantum $\Omega SU(k)$ sigma model. 
\\
\mbox{}\par\nobreak
\noindent
\textit{The Double Loop Algebra and Loop Algebra in our Sigma Model} 

The main aim of this subsection is to show that the conserved Noether currents corresponding to the $LSU(k)$ symmetry of $\Omega SU(k)$ generate the double loop algebra $LL\mathfrak{su}(k)$ in the quasi-topological model, which reduces to the loop algebra $L\mathfrak{su}(k)$ in the topological model. This shall be achieved by computing correlation functions of the Noether current with itself and with the energy-momentum tensor in the quasi-topological model, and taking the topological limit at the end.

Recall from the discussion below \eqref{locaction2} that correlation functions of  observables in the $\overline{Q}_+$-cohomology can be expressed as
\begin{equation}\label{corrfunc}
\langle\prod_{\gamma} \widetilde{\mathcal{O}}_{\gamma}\rangle=\sum_N e^{-tN}\int_{F_N}\mathcal{D}\phi\mathcal{D}\overline{\phi}\mathcal{D}\rho_{\overline{z}}\mathcal{D}\overline{\rho}_z\mathcal{D}\chi\mathcal{D}\overline{\chi}e^{-S_{pert.}(t)} \prod_{\gamma}\widetilde{\mathcal{O}}_{\gamma},
\end{equation}
where $S_{pert.}$ was defined in \eqref{action}.
Since only $S_{pert.}$ appears in \eqref{corrfunc}, one only needs to pay attention to this part of the action when computing the Noether current or correlation functions. Physically, this corresponds to performing perturbation theory around each vacuum labelled by $N$.

An action with {\it off-shell} supersymmetry is required in order to compute a correlation function. To this end, we shall consider \cite{Witten1988} the action\footnote{This action is equation 2.14 in \cite{Witten1988}, modulo the topological term, and with the field redefinition $H_{zam}=p_{zam}-\Gamma^{dq}_{am,cp}\rho_{zdq}\chi^{cp}$.} 
\begin{equation}\label{equivalentaction}
\begin{aligned}
S_{equiv}=&\int d^2z \Big(  p_{zam}\partial_{\overline{z}}\phi^{am}+\overline{p}_{\overline{z}am}\partial_{z}\overline{\phi}^{am}+\rho_{zam}\partial_{\overline{z}}\chi^{am}+\overline{\rho}_{\overline{z}am}\partial_{z}\overline{\chi}^{am}\\
&-t^{-1}g^{bn,am}(p_{zam}-\Gamma^{dq}_{am,cp}\rho_{zdq}\chi^{cp})(\overline{p}_{\overline{z}bn}-\overline{\Gamma}^{ek}_{bn,hl}\overline{\rho}_{\overline{z}ek}\overline{\chi}^{hl})
\\&-t^{-1}g^{am,hl} R^{ek}_{\textrm{     }bn,dq,am}\rho_{zek}\overline{\rho}_{\overline{z}hl}\chi^{bn}\overline{\chi}^{dq} \Big)+t\int \Phi^*\omega\\
=&S_{pert.}(t) + t \int \Phi^*\omega
\end{aligned}
\end{equation}
Here, $\rho_{zam}=t g_{am,bn}\overline{\rho}_{z}^{bn}$ and $\overline{\rho}_{\overline{z}am}=t g_{am,bn}\rho_{\overline{z}}^{bn}$, i.e., $\rho_{zam}\in \Gamma(K \otimes \Phi^*{T^*\Omega SU(k)})$ and $\overline{\rho}_{\overline{z}am}\in \Gamma(\overline{K} \otimes \Phi^*\overline{T^*\Omega SU(k)})$. From $S_{equiv}$ above, the algebraic equations of motion for the auxiliary fields $p_{zam}$ and $\overline{p}_{\overline{z}am}$ are given by
\begin{equation}\label{peom}
\begin{aligned}
p_{zam}&=t g_{am,bn}\partial_{z}\overline{\phi}^{bn}+\Gamma^{dq}_{am,cp}\rho_{zdq}\chi^{cp},\\
\overline{p}_{\overline{z}bn}&=t g_{am,bn}\partial_{\overline{z}}\phi^{am}+\overline{\Gamma}^{ek}_{bn,hl}\overline{\rho}_{\overline{z}ek}\overline{\chi}^{hl}.
\end{aligned}
\end{equation}
When the above explicit expressions of $p_{zam}$ and $\overline{p}_{\overline{z}bn}$ are substituted back into \eqref{equivalentaction}, one obtains \eqref{action}. In other words, $S$ and $S_{equiv}$ define the same theory. 

The supersymmetry transformations generated by $Q_-$ and $\overline{Q}_+$ now take a simple form:
\begin{equation}
\begin{aligned}[c]
\delta \phi^{am} &= \epsilon_+\chi^{am},\\
\delta \rho_{zam} &=-\epsilon_+p_{zam},\\
\delta\chi^{am}   &= 0,\\
\delta p_{zam} &= 0,
\end{aligned}
\\
\begin{aligned}[c]
\textrm{   }\\
\textrm{   }\\
\end{aligned}
\\
\begin{aligned}[c]
\delta \overline{\phi}^{am} &= \overline{\epsilon}_-\overline{\chi}^{am},\\
\delta \overline{\rho}_{\overline{z}am} &=-\overline{\epsilon}_-\overline{p}_{\overline{z}am},\\
\delta\overline{\chi}^{am}   &= 0,\\
\delta \overline{p}_{\overline{z}am} &= 0,
\end{aligned}
\end{equation}
and the action \eqref{equivalentaction} is invariant under these transformations, which satisfy $\delta^2=0$ without using the equations of motion.

Before proceeding to calculate the Noether current for the global $LSU(k)$ symmetry of the action, let us note that $\rho_{zam}$ and $\overline{\rho}_{\overline{z}am}$ transform under coordinate reparametrizations on $\Omega SU(k)$ as the components of (1,0) and (0,1) forms, respectively, and not as tangent vectors. Meanwhile, $p_{zam}$ and $\overline{p}_{\overline{z}am}$ are the components of one-forms on the worldsheet, but have complicated non-tensorial transformations under coordinate reparametrizations of $\Omega SU(k)$, as one may infer by inspecting their equations of motion \eqref{peom}. Such coordinate reparametrizations include the $LSU(k)$ isometry. Luckily, we do not need the precise transformations of $\rho_{zam}$, $\overline{\rho}_{\overline{z}am}$, $p_{zam}$ and $\overline{p}_{\overline{z}am}$ under the global $LSU(k)$ symmetry to compute the Noether current, since the derivatives of these fields do not appear in \eqref{equivalentaction}.

Following the arguments surrounding equation \eqref{corrfunc}, we only need to consider the pertubative part, $S_{pert.}(t)$, of the off-shell supersymmetric action \eqref{equivalentaction}. With the corresponding Lagrangian density denoted as $\mathcal{L}$, the standard formula 
\begin{equation}
\sum_{k\in \mathbb{Z}} \widetilde{J}^{\mu bk} \zeta_{b}^{k}=\dfrac{\partial \mathcal{L}}{\partial(\partial_\mu \phi^{an})}\delta \phi^{an} + \dfrac{\partial \mathcal{L}}{\partial(\partial_\mu \overline{\phi}^{an})}\delta \overline{\phi}^{an} +\dfrac{\partial \mathcal{L}}{\partial(\partial_\mu \chi^{an})}\delta \chi^{an}+\dfrac{\partial \mathcal{L}}{\partial(\partial_\mu \overline{\chi}^{an})}\delta \overline{\chi}^{an},  
\end{equation}
gives us the current, whose components are:
\begin{equation}\label{currents}
\begin{aligned}
\widetilde{J}^{bk}_z&=\frac{1}{2}if^{ab}_c( p_{zam}  \phi^{c\{m-k\}}+ \rho_{zam}  \chi^{c\{m-k\}}),\\
\widetilde{J}^{bk}_{\overline{z}}&=\frac{1}{2}if^{ab}_c( \overline{p}_{\overline{z}am}  \overline{\phi}^{c\{m-k\}}+ \overline{\rho}_{\overline{z}am}  \overline{\chi}^{c\{m-k\}}),
\end{aligned}
\end{equation}
where we have used $\eta^{z\overline{z}}=2$. We shall derive the current algebra using the $\widetilde{J}_z^{bk}$ component, rescaled as
\begin{equation}\label{rescaledJ}
2\widetilde{J}_z^{bk}=J_z^{bk}=if^{ab}_c( p_{zam}  \phi^{c\{m-k\}}+ \rho_{zam}  \chi^{c\{m-k\}}).
\end{equation} 
It is crucial to note that $J_z^{bk}$ is both $Q_A$-invariant and $\overline{Q}_+$-invariant, off-shell. Next, note that the holomorphic component of the energy-momentum tensor is 
\begin{equation}\begin{aligned}
T_{zz}&=\frac{1}{2}\Big(\frac{\partial\mathcal{L}}{\partial(\partial_{\overline{z}}\phi^{am})}\partial_{z}\phi^{am}+\frac{\partial\mathcal{L}}{\partial(\partial_{\overline{z}}\chi^{am})}\partial_{z}\chi^{am}\Big)
&=\frac{1}{2}(p_{zam}\partial_{z}\phi^{am}+\rho_{zam}\partial_{z}\chi^{am})
\end{aligned},
\end{equation}
and is also $Q_A$-invariant and $\overline{Q}_+$-invariant, off-shell. 

We would now like to compute the correlation functions $\langle J_z^{ak}(z,\overline{z})J_z^{bl}(w,\overline{w})\rangle$ and $\langle T_{zz}(z)J_z^{bl}(w,\overline{w})\rangle$. However, one should first note that $J_z^{bk}$ and $T_{zz}$ are in fact $Q_A$-{\it exact}, due to their nonzero holomorphic conformal dimensions, according to arguments below \eqref{confdime}. Consequently, the correlation functions we wish to calculate would vanish in the topological model. Hence, we shall calculate these correlation functions solely in the quasi-topological model.

A priori, the OPEs between the fundamental fields of the theory are complicated, and it is difficult to compute the correlation functions we want. To overcome this, we shall take the weak-coupling or infinite-volume limit of the target space, $\Omega SU(k)$, which corresponds to taking the limit where $t\rightarrow \infty$ in $S_{pert.}(t)$.
To understand why we are able to take the infinite-volume limit in our quasi-topological sigma model, note that 
the pertubative action can be written as
\begin{equation}\label{complexstructuredependence}
S_{pert.}(t)=\int d^2z (\{\overline{Q}_+,W'(t)\}+p_{zam}\partial_{\overline{z}}\phi^{am}+\rho_{zam}\partial_{\overline{z}}\chi^{am}),
\end{equation}
with
\begin{equation}
W'(t)=-\overline{\rho}_{\overline{z}am}\partial_z\overline{\phi}^{am}-t^{-1}g^{am,bn}\Gamma^{dk}_{cp,bn} \rho_{zdk}\chi^{cp}\overline{\rho}_{\overline{z}am}+t^{-1}g^{am,bn}p_{zam}\overline{\rho}_{\overline{z}bn},
\end{equation}
where \eqref{complexstructuredependence} consists of a $\overline{Q}_+$-exact term, and terms which depend solely on the complex structure of $\Omega SU(k)$. Correlation functions of $\overline{Q}_+$- closed observables are then independent of the value of $t$, as shown in \eqref{tindep}.


In particular, the correlation functions  $\langle J_z^{ak}(z,\overline{z})J_z^{bl}(w,\overline{w})\rangle$ and $\langle T_{zz}(z)J_z^{bl}(w,\overline{w})\rangle$ are preserved when taking the infinite-volume limit, since both $J^{bk}_z$ and $T_{zz}$ are $\overline{Q}_+$- closed observables. 
Taking this limit for \eqref{complexstructuredependence}, we obtain: 
\begin{equation}
S_{weak}=\int d^2z \Big(  p_{zam}\partial_{\overline{z}}\phi^{am}+\overline{p}_{\overline{z}am}\partial_{z}\overline{\phi}^{am}+\rho_{zam}\partial_{\overline{z}}\chi^{am}+\overline{\rho}_{\overline{z}am}\partial_{z}\overline{\chi}^{am}) 
\end{equation}
We find that we have an infinite number of $bc-\beta\gamma$ systems. The supersymmetries, the form of $J_z^{bk}$ and the form of $T_{zz}$ remain the same. 
The equations of motion in the infinite-volume limit indicate that $p_{zam}$, $\rho_{zam}$, $\phi^{am}$ and $\chi^{am}$ are holomorphic and $\overline{p}_{\overline{z}am}$, $\overline{\rho}_{\overline{z}am}$, $\overline{\phi}^{am}$ and $\overline{\chi}^{am}$ are antiholomorphic. Consequently, $J^{bk}_z$ is holomorphic.
The OPEs between the fields are
\begin{equation}
\begin{aligned}[c]
p_{zam}(z)\phi^{bn}(w)&\sim -\frac{\delta^{b}_{a}\delta^{n}_{m}}{z-w},\\
\rho_{zam}(z)\chi^{bn}(w)&\sim \frac{\delta^{b}_{a}\delta^{n}_{m}}{z-w},\\
\end{aligned}
\\
\begin{aligned}[c]
\textrm{   }\\
\textrm{   }\\
\end{aligned}
\\
\begin{aligned}[c]
\textrm{   }\\
\textrm{   }\\
\end{aligned}
\\
\begin{aligned}[c]
\overline{p}_{\overline{z}am}(\overline{z})\overline{\phi}^{bn}(\overline{w})&\sim -\frac{\delta^{b}_{a}\delta^{n}_{m}}{\overline{z}-\overline{w}},\\
\overline{\rho}_{\overline{z}am}(\overline{z})\overline{\chi}^{bn}(\overline{w})&\sim \frac{\delta^{b}_{a}\delta^{n}_{m}}{\overline{z}-\overline{w}}.\\
\end{aligned}
\end{equation}

Using these relations, we arrive at 
\begin{equation}\label{OPEcurrent}
J_z^{an_1}(z)J_z^{bn_2}(w)\sim \frac{if^{ab}_c J_z^{c\{n_1+n_2\}}(w)}{z-w},
\end{equation}
and
\begin{equation}\label{TJOPE}
T_{zz}(z)J_z^{ak}(w) \sim  \frac{J_z^{ak}(w)}{(z-w)^2}+\frac{\partial J_z^{ak}(w)}{(z-w)}.
\end{equation}
Using the Laurent expansions
\begin{equation}
J^{an}_z (z)=\sum_{m\in\mathbb{Z}}z^{-m-1}J^{an}_m
\end{equation} 
and 
\begin{equation}
T_{zz} (z)=\sum_{m\in\mathbb{Z}}z^{-m-2}L_m,
\end{equation} 
where 
\begin{equation}
J^{an}_{m}=\frac{1}{2\pi i}\oint dz z^m J^{an}_{z}(z)
\end{equation}
and
\begin{equation}
L_m=\frac{1}{2\pi i}\oint dz z^{m+1} T_{zz}(z),
\end{equation}
and the relation between operator commutators and their OPEs
\begin{equation}\label{commu}
[A,B]=\oint_0 dw \oint_w dz\textrm{ }a(z)b(w),
\end{equation}
where $A=\oint a(z)dz$ and $B=\oint b(w)dw$ are operators while $a(z), b(w)$ are holomorphic fields, we find that \eqref{OPEcurrent} and \eqref{TJOPE} respectively imply the double loop algebra $LL\mathfrak{g}$
\begin{equation}\label{doubleloop}
[J^{an_1}_{m_1},J^{bn_2}_{m_2}]=if^{ab}_cJ^{c\{n_1+n_2\}}_{m_1+m_2},
\end{equation}
and 
\begin{equation}\label{energyraising}
[L_n,J^{ak}_m]=-mJ^{ak}_{n+m}.
\end{equation}

In particular, we have
\begin{equation}\label{energyraising2}
[L_0,J^{ak}_m]=-mJ^{ak}_{m},
\end{equation}
i.e., the current algebra is the spectrum-generating algebra of our quasi-topological model.\footnote{Recall from the discussion below \eqref{locaction2} that our model only has at most excited states with antiholomorphic conformal dimension equal to zero, i.e., $\overline{L}_0$ always has eigenvalue zero.} The quantum topological model forms a subsector of the quasi-topological model, and consists solely of ground states. As such, \eqref{energyraising2} implies that the only current algebra generators that can act on elements of the Hilbert space of the topological model are $J^{ak}_0$, which generate the loop algebra $L\mathfrak{su}(k)$, that is an affine Lie algebra with no central extension:
\begin{equation}\label{singleloop}
[J^{an_1}_{0},J^{bn_2}_{0}]=if^{ab}_cJ^{c\{n_1+n_2\}}_{0}.
\end{equation}
Therefore, the double loop current algebra $LL\mathfrak{su}(k)$ effectively becomes a (single) loop current algebra $L\mathfrak{su}(k)$ in the topological sigma model. In short, the double loop algebra $LL\mathfrak{su}(k)$ appears in the quasi-topological sigma model while the loop algebra $L\mathfrak{su}(k)$ appears in the topological sigma model.

\mbox{}\par\nobreak
\noindent
\textit{The Appearance of Central Extensions}

One always obtains projective representations of symmetry groups in quantum theories, since a state $|\alpha\rangle$ which represents a quantum system is equivalent to the state $e^{i\nu}|\alpha\rangle$, where $\nu$ is a phase (\cite{gannon2006moonshine}, Chapter 3). It is known that projective representations of the loop group $LG$ lift to representations of central extensions of $LG$ \cite{pressley1986loop}. Hence, since the double loop algebra $LL\mathfrak{su}(k)$ \eqref{doubleloop} essentially contains two copies of the loop algebra $L\mathfrak{su}(k)$ (which one can see by setting $(n_1,n_2)=(0,0)$ or $(m_1,m_2)=(0,0)$), at the quantum level, we should obtain projective representations of both copies of the loop algebra, which each lift to representations of central extensions of the loop algebras themselves. 

In our model, we may understand the appearance of these central extensions as being due to a quantum anomaly of the classical $LSU(k)$ symmetry of our model. Our derivation of the conserved Noether current $J_z^{bk}$ was derived from a classical Lagrangian density; hence, the expression for $J_z^{bk}$ is valid only when the $LSU(k)$ symmetry is not anomalous, i.e., when the path integral measure is invariant under the symmetry transformations. However, this is not necessarily true, and the aforementioned central extensions can be considered to be quantum corrections due to an anomaly. In other words, we can associate to our quasi-topological sigma model a toroidal Lie algebra $\mathfrak{su}(k)_\textrm{tor}$:
\begin{equation}\label{toroidal}
[J^{a n_1}_{m_1},J^{b n_2}_{m_2}]=if^{ab}_cJ^{c\{n_1+n_2\}}_{m_1+m_2}+c_1 n_1\delta^{ab} \delta^{\{n_1+n_2\}0} \delta_{\{m_1+m_2\}0} +c_2 m_1\delta^{ab} \delta^{\{n_1+n_2\}0} \delta_{\{m_1+m_2\}0},
\end{equation}
and to the topological sigma model an affine Lie algebra $\mathfrak{su}(k)_\textrm{aff}$: 
\begin{equation}\label{affinecurrent}
[J^{a n_1}_0,J^{b n_2}_0]=if^{ab}_cJ^{c\{n_1+n_2\}}_0+c_1 n_1\delta^{ab} \delta^{\{n_1+n_2\}0}.
\end{equation}

\section{The $A_{k-1}$ $\mathcal{N}=(2,0)$ Little String Theory}
\subsection{Mapping Local Observables via Atiyah's Theorem}

Recall that the DLCQ of the $A_{k-1}$ $\mathcal{N}=(2,0)$ LST on $S^1 \times \R\times \R^4 $ with $N$ units of momentum is given by the two-dimensional $\mathcal{N} = (4,4)$ supersymmetric sigma model with $\mathcal{M}^N_{SU(k)}(\R^4)$  target space.\footnote{
Since the NS5-brane worldvolume theory is physically sensible and unitary, one may analytically continue the Lorentzian worlvolume $(S^1\times \R)^{1,1}\times \R^4$ to the Euclidean worldvolume $S^1 \times \R\times \R^4 $. It will be useful for our purposes to adopt the Euclidean signature. }
The Hilbert space of ground states of this sigma model corresponds to its topological sector, and the further inclusion of left-excited states gives us its quasi-topological sector.\footnote{
We concentrate on the pertubative sector of the $\mathcal{N}$=(4,4) sigma model and exclude worldsheet instantons, because the target space $\mathcal{M}_{SU(k)}^N(\R^4)$ is  hyperk{\"a}hler and its quantum cohomology reduces to ordinary cohomology  \cite{DijkgraafInstantonStrings}.}

The ground states of the $\mathcal{N}$=(4,4) sigma model with target space $\mathcal{M}_{SU(k)}^N(\R^4)$ correspond to the $L^2$-cohomology of $\mathcal{M}_{SU(k)}^N(\R^4)$ as local observables {\cite{Witten1988}. 
This is equivalent, via Atiyah's theorem, to the $L^2$-cohomology of $\mathcal{M}(\C P^1 \xrightarrow[hol.]{N}\Omega{SU(k)})$,\footnote{The $L^2$-cohomology consists of topological invariants which are preserved by the diffeomorphism between $\mathcal{M}_{SU(k)}^N(\R^4)$ and $\mathcal{M}(\C P^1 \xrightarrow[hol.]{N}\Omega{SU(k)})$.} which are the local observables of 
the nonpertubative topological $\Omega SU(k)$ sigma model.

 Likewise, the left-excited states of the $\mathcal{N}$=(4,4) sigma model with target space $\mathcal{M}_{SU(k)}^N(\R^4)$ correspond to local observables described by the \v{C}ech cohomology of the sheaf of chiral de Rham complex on $\mathcal{M}_{SU(k)}^N(\R^4)$
{\cite{kapustin2005chiral,Tan2006}}.
This is equivalent, via Atiyah's theorem, to the \v{C}ech cohomology of the sheaf of chiral de Rham complex on $\mathcal{M}(\C P^1 \xrightarrow[hol.]{N}\Omega{SU(k)})$,\footnote{\label{sheafdiff}
Sections of the sheaf of Chiral de Rham complex are invariant under diffeomorphisms, since diffeomorphisms are geometrical automorphisms of the theory described by the Chiral de Rham complex \cite{Tan2006}.}
} which are the local observables of 
the quasi-topological sector of the auxiliary theory defined in Section 3.1, which is an $\mathcal{N}=(2,2)$ sigma model with target space $\mathcal{M}(\C P^1 \xrightarrow[hol.]{N}\Omega{SU(k)})$ that is associated with the $\Omega SU(k)$ sigma model.

In this way, by studying the topological and quasi-topological $\Omega SU(k)$ sigma models, one can analyze the ground and left-excited states of the $A_{k-1}$ $\mathcal{N}=(2,0)$ little string theory.
\subsection{Local Observables as Modules over the Current Algebra}

We shall now exploit the above relations 
to show that the ground and left-excited states of the $A_{k-1}$ $\mathcal{N}=(2,0)$ little string theory form modules over the affine Lie algebra $\mathfrak{su}(k)_{\textrm{aff}}$ and toroidal Lie algebra $\mathfrak{su}(k)_{\textrm{tor}}$.

\mbox{}\par\nobreak
\noindent
\textit{Module over Affine Lie Algebra} 

Recall that the current $J_z^{bk}(z)$ given in \eqref{rescaledJ} is a $\overline{Q}_+$-closed operator, which is not $\overline{Q}_+$-exact. When we Laurent expand $J_z^{bk}(z)$, we find that these properties are inherited by all its Laurent modes, which, as we have shown, generate $\mathfrak{su}(k)_{\textrm{tor}}$. Also recall that $J_z^{bk}(z)$ is $Q_A$-exact, due to a nonzero holomorphic conformal dimension. However, observe that the Laurent zero modes $J_0^{bk}$ which generate $\mathfrak{su}(k)_{\textrm{aff}}$ in the topological model have holomorphic conformal dimension equal to zero, and as such, cannot become $Q_A$-exact using arguments below \eqref{confdime}. They are in fact $Q_A$-closed, since
\begin{equation}\label{qaclosedjo}
[Q_A,J^{am}_0]=\frac{1}{2\pi i}\oint dz [Q_A,J^{am}_z (z)]=0.
\end{equation}   

We previously found the affine Lie algebra $\mathfrak{su}(k)_{\textrm{aff}}$ as a current algebra in the topological $\Omega SU(k)$ sigma model. Starting with a ground state $|0\rangle$ of this theory, which should be $Q_A$-closed ($Q_A|0\rangle$=0)
, we can act with generators of the affine Lie algebra $\mathfrak{su}(k)_{\textrm{aff}}$ to obtain other states in the theory in the form of a highest weight module over $\mathfrak{su}(k)_{\textrm{aff}}$, i.e.,
\begin{equation}\label{module1}
J_0^{a\{-n_1\}} J_0^{b\{-n_2\}} J_0^{c\{-n_3\}}\ldots|0 \rangle ,
\end{equation}
where $n_i\geq 0$.\footnote{The Sugawara construction can be used on the affine Lie algebra generators to obtain a grading operator $\widehat{L}_0$, which acts on \eqref{module1} to give the eigenvalue $\varepsilon-n_1-n_2-n_3\ldots$, where $\varepsilon$ is the grade of $|0\rangle$. To be a well-defined highest weight state, $|0\rangle$ must be annihilated by $J^{-n_i}_0$ with $n_i< 0$.}
These states are $Q_A$-closed due to \eqref{qaclosedjo}, e.g., for a state $J^{a\{-k\}}_0|0\rangle$,
\begin{equation}\label{Qcohom1}
Q_AJ^{a\{-k\}}_0|0\rangle = [Q_A, J^{a\{-k\}}_0]|0\rangle =0,
\end{equation}
and in a similar manner, $Q_A$ annihilates the other states of the form \eqref{module1} due to the $Q_A$-invariance of all the affine Lie algebra generators $J^{bk}_0$.
In addition, it is not possible for any of the states \eqref{module1} to be $Q_A$-exact. This can be explained as follows. If any such state was $Q_A$-exact, then we would have 
\begin{equation}\label{Qcohom2}
J_0^{a\{-n_1\}} J_0^{b\{-n_2\}} J_0^{c\{-n_3\}}\ldots|0 \rangle  = Q_A|\Psi\rangle = [Q_A, \Psi]|0\rangle,
\end{equation}
where $\Psi$ is some operator giving rise to the corresponding state $|\Psi\rangle$ by acting on $|0\rangle$.
However, $J_0^{ak}$ is not a $Q_A$-exact operator, whence it follows that a product of affine Lie algebra generators also cannot be a $Q_A$-exact operator; in turn, this means that we cannot have \eqref{Qcohom2}, and thus the states \eqref{module1} cannot be $Q_A$-exact. Therefore, since the states \eqref{module1} are $Q_A$-closed but not $Q_A$-exact, they are elements of the $Q_A$-cohomology. As the generators of the affine Lie algebra $\mathfrak{su}(k)_{\textrm{aff}}$ do not raise the energy level of the states (according to \eqref{energyraising2}), all the states of the form \eqref{module1} are sigma model ground states. 

As noted in the discussion below \eqref{BPS}, the $Q_A$-cohomology of ground states in each worldsheet instanton sector corresponds to local observables which can be identified as elements of the $L^2$-cohomology of $\mathcal{M}(\C P^1 \xrightarrow[hol.]{N}\Omega{SU(k)})$, which is finite-dimensional. This must mean that that the $Q_A$-cohomology of ground states in a particular worldsheet instanton sector $N$ should consist of a finite-dimensional submodule $\widehat{su}(k)^N_{c_1}$ over $\mathfrak{su}(k)_{\textrm{aff}}$.
In other words, the $L^2$-cohomology of $\mathcal{M}(\C P^1 \xrightarrow[hol.]{N}\Omega{SU(k)})$ 
forms a finite submodule $\widehat{su}(k)^N_{c_1}$ over the affine Lie algebra $\mathfrak{su}(k)_{\textrm{aff}}$ (given in \eqref{affinecurrent}) of level $c_1$:
\begin{equation}
H_{L^2}^*(\mathcal{M}(\C P^1 \xrightarrow[hol.]{N}\Omega{SU(k)}))=\widehat{su}(k)_{c_1}^{N}.
\end{equation}

Note that the $L^2$-cohomology of $\mathcal{M}(\C P^1 \xrightarrow[hol.]{N}\Omega{SU(k)})$ is isomorphic to its intersection cohomology, $\textrm{IH}^*(\mathcal{M}(\C P^1 \xrightarrow[hol.]{N}\Omega{SU(k)}))$ \cite{Goresky}. 
Then, via the diffeomorphism of Atiyah's theorem, we find that
\begin{equation}\label{bravefinkel}
\textrm{IH}^*(\mathcal{M}^N_{SU(k)}(\R^4))=\widehat{su}(k)_{c_1}^{N},
\end{equation}
since intersection cohomology consists of topological invariants. 

The space $\mathcal{M}^N_{SU(k)}(\R^4)$ admits a decomposition into subspaces of smaller dimension, since the instanton number $N$ can be decomposed as \cite{Braverman2010} 
\begin{equation}\label{instadecomp}
N=(i-j)+\frac{1}{2}(\overline{\lambda},\overline{\lambda})-\frac{1}{2}(\overline{\mu},\overline{\mu}),
\end{equation}
where $i$ and $j$ are integers satisfying $i\geq j$, whilst $\overline{\lambda}$ and $\overline{\mu}$ are vectors with dimension equal to the rank of $SU(k)$ and norm valued in $2\mathbb{N}$.\footnote{Although this decomposition of $N$ is known from the mathematical literature, it can be understood from the point of view of the  NS5-brane, by lifting to M-theory \cite{Tan2013a}. This is carried out by decompactifying the eleventh dimension, whereby the stack of NS5-branes becomes a stack of M5-branes. Then, identifying the circle, $S^1$, in the M5-brane worldvolume, $S^1\times \mathbb{R}\times\mathbb{R}^4$, as the eleventh dimension of M-theory, one obtains type IIA string theory upon compactification on $S^1$. Consequently, the stack of $k$ M5-branes reduces to a stack of $k$ D4-branes, and D0-branes appear as Kaluza-Klein modes in the D4-brane worldvolume, $\mathbb{R} \times\mathbb{R}^4$. 
Bound states of these D0-branes give rise to static, particle-like BPS configurations in the D4-brane worldvolume, which in turn appear as $SU(k)$ instantons on $\R^4$. 
In \eqref{instadecomp}, $(i-j)$ counts the total number of D0-branes in $\R^4/\{0\}$ which contribute to the bound state, $\frac{1}{2}(\overline{\lambda},\overline{\lambda})$ gives the number of D0-branes at the origin of $\R^4$ which contribute, while the subtraction by $\frac{1}{2}(\overline{\mu},\overline{\mu})$ accounts for D0-branes at infinity, which contribute instanton number zero since they necessarily correspond to flat gauge fields in order for the action of the instanton to be finite (\cite{Vafa1994}, Section 4.4).} The vectors $\overline{\lambda}$ and $\overline{\mu}$ are required to be dominant coweights of $SU(k)$ \cite{Tan2013a}, which are identified with dominant weights of the Langlands dual of $SU(k)$. Furthermore, $\overline{\lambda}$ and $\overline{\mu}$ can also be regarded as part of the triples $\widehat{\lambda}=(\overline{\lambda},1,i)$ and $\widehat{\mu}=(\overline{\mu},1,j)$. Since $\widehat{\lambda}$ and $\widehat{\mu}$ completely specify the instanton number \eqref{instadecomp}, we have the decomposition
\begin{equation}\label{modspacedecomp}
\textrm{IH}^*(\mathcal{M}^N_{SU(k)}(\R^4))=\bigoplus_{\widehat{\lambda},\widehat{\mu}}\textrm{IH}^*(\mathcal{M}^{\widehat{\lambda},\widehat{\mu}}_{SU(k)}(\R^4)).
\end{equation}


This decomposition of the instanton moduli space, together with \eqref{bravefinkel}, mean that the affine submodule in each worldsheet instanton sector $N$ should also decompose in an identical manner, i.e.,
\begin{equation}\label{affinedecomp}
\widehat{su}(k)_{c_1}^{N}=\bigoplus_{\widehat{\lambda},\widehat{\mu}}\widehat{su}(k)_{c_1}^{\widehat{\lambda},\widehat{\mu}}.
\end{equation}
We would like to ascertain the meaning of $\widehat{\lambda}$ and $\widehat{\mu}$ in the decomposition \eqref{affinedecomp}. Note that the triples that specify the decomposition \eqref{modspacedecomp} can be regarded as dominant coweights of the affine Kac-Moody group $SU(k)_\textrm{aff}$ of level 1 \cite{Tan2013a}, which are just dominant weights for the level 1 Langlands dual affine Kac-Moody group, $SU(k)^L_\textrm{aff}$. The Lie algebra of this group is just $\mathfrak{su}(k)_{\textrm{aff}}$. Thus, for level $c_1=1$, $\widehat{\lambda}$ and $\widehat{\mu}$ are dominant affine weights in the weight spaces of modules over $\mathfrak{su}(k)_{\textrm{aff}}$, and we have
\begin{equation}\label{bravefinkel2}
\textrm{IH}^*(\mathcal{M}^{\widehat{\lambda},\widehat{\mu}}_{SU(k)}(\R^4))=\widehat{su}(k)_{1}^{\widehat{\lambda},\widehat{\mu}},
\end{equation}
which is just Braverman and Finkelberg's result \cite{Braverman2010} for the case of $SU(k)$ instantons on $\R^4$. In other words, we have a sigma model derivation of their result.

\mbox{}\par\nobreak
\noindent
\textit{Module over Toroidal Lie Algebra} 

For the quasi-topological $\Omega SU(k)$ sigma model, we may follow a similar line of argument as that presented above, with $\overline{Q}_+$ instead of $Q_A$, the toroidal Lie algebra $\mathfrak{su}(k)_{\textrm{tor}}$ instead of the affine Lie algebra $\mathfrak{su}(k)_{\textrm{aff}}$, and chiral differential operators instead of $L^2$-cohomology. Now, acting on a ground state $|0\rangle$ (which is $\overline{Q}_+$-closed) with the generators of $\mathfrak{su}(k)_{\textrm{tor}}$, we can have excited states in addition to the ground states, i.e., the states are of the form 
\begin{equation}\label{module2}
J_{-m_1}^{a\{-n_1\}} J_{-m_2}^{b\{-n_2\}} J_{-m_3}^{c\{-n_3\}}\ldots|0 \rangle,
\end{equation}
where the Laurent indices $m_i\geq 0$. Note that in order to have a well-defined sigma model vacuum, $|0\rangle$ is annihilated by toroidal Lie algebra generators with $m_i < 0$. These states are excited for $m_i > 0$, i.e., they can have nonzero holomorphic conformal dimension (according to \eqref{energyraising2}); and can be shown to be elements of the $\overline{Q}_+$-cohomology using arguments analogous to those surrounding equations \eqref{Qcohom1} and \eqref{Qcohom2}. 
In other words, the Hilbert space of left-excited states  of the quasi-topological $\Omega SU(k)$ sigma model consists of elements of modules over the toroidal Lie algebra $\mathfrak{su}(k)_{\textrm{tor}}$.

In particular, for each worldsheet instanton sector $N$, the quasi-topological states will be left-excitations of the sigma model ground states in the finite affine submodule $\widehat{su}(k)_{c_1}^{N}$, and as such can be considered to be a submodule over the toroidal Lie algebra $\mathfrak{su}(k)_{\textrm{tor}}$. The toroidal Lie algebra submodule which contains the affine submodule $\widehat{su}(k)_{c_1}^{N}$ will be denoted $\widehat{\widehat{su}}(k)_{c_1,c_2}^{N}$.

As noted in the discussion below \eqref{locaction2}, the $\overline{Q}_+$-cohomology of states in this $N$-sector corresponds to local observables which can be identified as elements of the \v{C}ech cohomology of
\begin{equation}
\widehat{\Omega}^{ch}_{\mathcal{M}(\C P^1 \xrightarrow[hol.]{N} \Omega{SU(k)})},
\end{equation}
the sheaf of chiral de Rham complex on $\mathcal{M}(\C P^1 \xrightarrow[hol.]{N}\Omega{SU(k)})$.
The above discussion then implies that this cohomology forms a module $\widehat{\widehat{su}}(k)^N_{c_1,c_2}$ over the toroidal Lie algebra $\mathfrak{su}(k)_{\textrm{tor}}$ (given in \eqref{toroidal}) of levels $c_1$ and $c_2$:
\begin{equation}
H_{\textrm{\v{C}ech}}^*(\widehat{\Omega}^{ch}_{\mathcal{M}(\C P^1 \xrightarrow[hol.]{N}\Omega{SU(k)})})=\widehat{\widehat{su}}(k)_{c_1,c_2}^{N}.
\end{equation}

Via the diffeomorphism of Atiyah's theorem (see footnote \ref{sheafdiff}), this is equivalent to 

\begin{equation}\label{generalizedBF}
H_{\textrm{\v{C}ech}}^*(\widehat{\Omega}^{ch}_{\mathcal{M}^N_{SU(k)}(\R^4)})=\widehat{\widehat{su}}(k)_{c_1,c_2}^{N}.
\end{equation}
Clearly, \eqref{generalizedBF}, which states that the \v{C}ech cohomology on the sheaf of chiral de Rham complex on the moduli space of $SU(k)$ $N$-instantons on $\R^4$ forms a submodule over the toroidal Lie algebra $\mathfrak{su}(k)_{\textrm{tor}}$, is a generalization of Braverman and Finkelberg's result for the case of $SU(k)$ instantons on $\R^4$. 

Braverman and Finkelberg's result is closely related to the celebrated AGT correspondence that relates \textit{equivariant} intersection cohomology of $\mathcal{M}_{SU(k)}^N(\R^4)$ and affine $W$-algebras. To be precise, the introduction of equivariance to the ordinary intersection cohomology of $\mathcal{M}_{SU(k)}^N(\R^4)$ corresponds to a quantum Drinfeld-Sokolov reduction of $\mathfrak{su}(k)_{\textrm{aff}}$ to its associated affine $W$-algebra. Our generalization \eqref{generalizedBF} of Braverman and Finkelberg's result then suggests that the \textit{equivariant} \v{C}ech cohomology on the sheaf of chiral de Rham complex on $\mathcal{M}_{SU(k)}^N(\R^4)$ would be mathematically related to 
a `toroidal' $W$-algebra obtained using an analog of the quantum Drinfeld-Sokolov reduction on $\mathfrak{su}(k)_{\textrm{tor}}$.

\section{The M5-brane Worldvolume Theory}

Dijkgraaf conjectured in  \cite{Dijkgraaf1998,Dijkgraaf1998long} that the $\mathcal{N}=(4,4)$ supersymmetric sigma model with target $\mathcal{M}_{SU(k)}^N(\R^4)$ also describes the DLCQ of the worldvolume theory of $k$ M5-branes on $S^1\times \R \times \R^4$ where little strings wrap $S^1$ with $N$ units of discrete momenta. In this section,  we shall attempt to use our results from the previous section to understand the 1/2 and 1/4 BPS sectors of the M5-brane worldvolume theory. Here, by 1/2 BPS (1/4 BPS) sector, we mean the sector of the theory which is invariant under half (quarter) of the sixteen worldvolume supersymmetries.
\subsection{Little Strings and the M5-brane}

Firstly, we shall elaborate on the 1/2 and 1/4 BPS sectors of the M5-brane worldvolume theory, and explain how they are captured by the $\Omega SU(k)$ 
sigma model studied in the previous section. The 6d $\mathcal{N}=(2,0)$ supersymmetry algebra of the M5-brane worldvolume theory on $S^1\times \R \times \R^4$ is \cite{Dijkgraaf1996,dijkgraaf1997bps}
\begin{equation}\label{worldvolumealgebra}
\begin{aligned}
\{\mathcal{Q}^{a\alpha},\mathcal{Q}^{b\beta}\}&=\epsilon^{\alpha\beta}(\mathcal{H}\textbf{1}^{ab}+\mathcal{P}^m_L\Gamma^{ab}_m),\\
\{\mathcal{Q}^{a\dot{\alpha}},\mathcal{Q}^{b\dot{\beta}}\}&=\epsilon^{\dot{\alpha}\dot{\beta}}(\mathcal{H}\textbf{1}^{ab}+\mathcal{P}^m_R\Gamma^{ab}_m),
\end{aligned}
\end{equation}
where $a,b=1,\ldots,4$ are chiral spinor indices for the $Spin(5)$ spatial rotation group on the worldvolume; $\alpha,\beta=1,2$ ($\dot{\alpha},\dot{\beta}=1,2$) are chiral (anti-chiral) spinor indices for the $Spin(4)$ R-symmetry group; $\epsilon^{\alpha\beta}$ is the Levi-Civita tensor; $\mathcal{H}$ is the Hamiltonian of the worldvolume theory; $\mathcal{P}^m_L=\mathcal{P}^m+\mathcal{W}^m$ and $\mathcal{P}^m_R=\mathcal{P}^m-\mathcal{W}^m$, where $\mathcal{P}^m$ are the worldvolume momenta and $\mathcal{W}^m$ are the worldvolume winding numbers (note that in the case at hand only the winding along $S^1$ is non-zero)
; and $\Gamma_m$ are gamma matrices corresponding to the $Spin(5)$ spatial rotation group.\footnote{We are using Hamiltonian notation, whereby only the $Spin(5)$ spatial rotation group of the worldvolume is manifest.} 
 From here, the 1/2 BPS sector of the worlvolume theory is defined to consist of states which satisfy the four independent relations
\begin{equation}\label{BPSstatecon}
\begin{aligned}[c]
\varepsilon^1_{a\alpha}\mathcal{Q}^{a\alpha}|BPS\rangle=0,
\end{aligned}
\\
\begin{aligned}[c]
\textrm{   }\\
\textrm{   }\\
\end{aligned}
\\
\begin{aligned}[c]
\varepsilon^2_{a\alpha}\mathcal{Q}^{a\alpha}|BPS\rangle=0,
\end{aligned}
\\
\begin{aligned}[c]
\textrm{   }\\
\textrm{   }\\
\end{aligned}
\\
\begin{aligned}[c]
\varepsilon^3_{a\alpha}\mathcal{Q}^{a\alpha}|BPS\rangle=0,
\end{aligned}
\\
\begin{aligned}[c]
\textrm{   }\\
\textrm{   }\\
\end{aligned}
\\
\begin{aligned}[c]
\varepsilon^4_{a\alpha}\mathcal{Q}^{a\alpha}|BPS\rangle=0,
\end{aligned}
\end{equation}
where $\varepsilon^i_{a\alpha}$  are four different chiral (with respect to the $Spin(4)$ R-symmetry) spinors; as well as the four independent relations
\begin{equation}\label{BPSstatecon2}
\begin{aligned}[c]
\varepsilon^1_{a\dot{\alpha}}\mathcal{Q}^{a\dot{\alpha}}|BPS\rangle=0,
\end{aligned}
\\
\begin{aligned}[c]
\textrm{   }\\
\textrm{   }\\
\end{aligned}
\\
\begin{aligned}[c]
\varepsilon^2_{a\dot{\alpha}}\mathcal{Q}^{a\dot{\alpha}}|BPS\rangle=0,
\end{aligned}
\\
\begin{aligned}[c]
\textrm{   }\\
\textrm{   }\\
\end{aligned}
\\
\begin{aligned}[c]
\varepsilon^3_{a\dot{\alpha}}\mathcal{Q}^{a\dot{\alpha}}|BPS\rangle=0,
\end{aligned}
\\
\begin{aligned}[c]
\textrm{   }\\
\textrm{   }\\
\end{aligned}
\\
\begin{aligned}[c]
\varepsilon^4_{a\dot{\alpha}}\mathcal{Q}^{a\dot{\alpha}}|BPS\rangle=0,
\end{aligned}
\end{equation}
where $\varepsilon^i_{a\dot{\alpha}}$  are four different anti-chiral spinors.
Now, note that the first set of relations involves only chiral supercharges and the second set of relations involves only anti-chiral supercharges. The 1/4 BPS sector which we consider corresponds to dropping the four chiral relations given in \eqref{BPSstatecon}. 

Since the little strings that live on the M5-brane worldvolume are 1/2 BPS objects, they only preserve eight of the sixteen supersymmetries of the worldvolume theory. The eight preserved supercharges on each little string worldsheet obey the supersymmetry algebra  \cite{Dijkgraaf1996,dijkgraaf1997bps}
\begin{equation}\label{littlealgebra}
\begin{aligned}
\{\mathscr{Q}^{\dot{a}\alpha},\mathscr{Q}^{\dot{b}\beta}\}&=2\epsilon^{\dot{a}\dot{b}}\epsilon^{\alpha\beta}\mathscr{L}_0,\\
\{\overline{\mathscr{Q}}^{\dot{a}\dot{\alpha}},\overline{\mathscr{Q}}^{\dot{b}\dot{\beta}}\}&=2\epsilon^{\dot{a}\dot{b}}\epsilon^{\dot{\alpha}\dot{\beta}}\overline{\mathscr{L}}_0,
\end{aligned}
\end{equation}
where $\dot{a},\dot{b}=1,2$ are anti-chiral spinor indices for the $Spin(4)$ rotation group of the $\R^4$ transverse to the worldsheet, and $\mathscr{L}_0$ and $\overline{\mathscr{L}}_0$ are the left and right-moving parts of the Hamiltonian on the worldsheet. Observe that the eight preserved supercharges are evenly divided into chiral ($\mathscr{Q}^{\dot{a}\alpha}$) and anti-chiral ($\overline{\mathscr{Q}}^{\dot{a}\dot{\alpha}}$) under the $Spin(4)$ R-symmetry of the worldvolume theory. Hence, the 1/2 BPS sector, defined by states obeying the eight chiral and anti-chiral relations \eqref{BPSstatecon} and \eqref{BPSstatecon2}, corresponds to states annihilated by all eight left-moving and right-moving supercharges on the worldsheet. From the supersymmetry algebra \eqref{littlealgebra}, we find that these states are necessarily ground states annihilated by the Hamiltonian $\mathscr{H}=\mathscr{L}_0+\overline{\mathscr{L}}_0$. Furthermore, the 1/4 BPS sector, which only obeys the anti-chiral relations \eqref{BPSstatecon2}, corresponds to states annihilated by the four right-moving supercharges on the worldsheet, and therefore, from \eqref{littlealgebra}, can be left-excited by $\mathscr{L}_0\neq0$. 

 According to Dijkgraaf
\cite{Dijkgraaf1998,Dijkgraaf1998long}, the worldvolume dynamics of a stack of $k$ M5-branes is captured by this little string theory, 
 whose DLCQ is described by the two-dimensional $\mathcal{N} = (4,4)$ supersymmetric sigma model with $\mathcal{M}^N_{SU(k)}(\R^4)$  target space. 

The Hilbert space of ground states of this sigma model corresponds to its topological sector, and the further inclusion of left-excited states gives us its quasi-topological sector. 
Consequently, the 1/2 (1/4) BPS sector of the worldvolume theory of a stack of $k$ M5-branes, with center-of-mass dynamics frozen, can be described by the topological (quasi-topological) sector of the $\mathcal{N}$=(4,4) sigma model on $S^1 \times \R $ with target space $\mathcal{M}_{SU(k)}^N(\R^4)$, 
for all $N>0$ \citep{Dijkgraaf1998,Dijkgraaf1998long}.
\footnote{
As in the previous section, we just need to concentrate on the pertubative sector of the $\mathcal{N}$=(4,4) sigma model, without worldsheet instantons.}
Only $N>0$ shall be considered in the following, 
since $N$ corresponds to the units of momenta along the lightcone circle.  Then, from Section 4.1, we find that the 1/2 BPS sector is given by the $L^2$-cohomology of $\mathcal{M}(\C P^1 \xrightarrow[hol.]{N}\Omega{SU(k)})$, 
 while the 1/4 BPS sector is given by \v{C}ech cohomology of the sheaf of chiral de Rham complex on $\mathcal{M}(\C P^1 \xrightarrow[hol.]{N}\Omega{SU(k)})$. 
Hence, by analyzing the topological and quasi-topological $\Omega SU(k)$ sigma models, one can understand the 1/2 and 1/4 BPS sectors of the M5-brane worldvolume theory.
Let us now compute the partition functions of the 1/2 BPS and 1/4 BPS sectors of the M5-brane. 
\subsection{1/2 BPS Sector}
The Hilbert space (in discrete lightcone gauge) of the worldvolume theory of a stack of $k$ M5-branes is
graded \cite{Dijkgraaf1998}, i.e., it is given as
\begin{equation}\label{gradedhilbert}
\mathcal{H}_p=\bigoplus_{N>0}p^N \mathcal{H}_N,
\end{equation}
where $p$ is a complex parameter, and where $\mathcal{H}_N$ is the Hilbert space of the two-dimensional $\mathcal{N}=(4,4)$ sigma model on the moduli space of $SU(k)$ $N$-instantons on $\R^4$. The 1/2 BPS sector of the worldvolume theory is given by the topological sector of \eqref{gradedhilbert}, which is captured by the topological sigma model with $\Omega SU(k)$ target space (for $N>0$), as explained in Section 5.1. In other words, the Hilbert space of our topological sigma model also has the graded structure given in \eqref{gradedhilbert}. In light of this fact,
the partition function of the topological sigma model,
\begin{equation}\label{pfHam}
Z_{(top)}=\textrm{Tr}_{\mathcal{H}^{(top)}}e^{-\beta H},
\end{equation}
can be written (for $N>0$) as 
\begin{equation}
\sum_{N>0} p^N\textrm{Tr}_{\mathcal{H}_N^{(top)}}e^{-\beta H}.
\end{equation}
Here, $\mathcal{H}^{(top)}$ is the Hilbert space of the entire topological sigma model, $\mathcal{H}_N^{(top)}$ is the Hilbert space of a particular worldsheet instanton sector of the topological sigma model, $H$ is the Hamiltonian, and $\beta$ is a real parameter.
Since the topological sector only contains ground states whose eigenvalues under $H$ are zero, we have 
\begin{equation}\label{topppf}
\begin{aligned}
Z_{(top)}&=\sum_{N > 0} p^N \textrm{Tr}_{\mathcal{H}_N^{(top)}} e^{-\beta H}\\
&=\sum_{N > 0}p^N \textrm{dim} \mathcal{H}^{(top)}_N \\
&=\textrm{Tr}_{\mathcal{H}^{(top)}}p^{\widehat{N}},
\end{aligned}
\end{equation}
where $\widehat{N}$ is the instanton number operator which has $N$ as its eigenvalue.

From Section 4.2, we know that the 
Hilbert space $\mathcal{H}^{(top)}$ is made up of submodules over the affine Lie algebra $\mathfrak{su}(k)_{\textrm{aff}}$, one for every worldsheet instanton sector $N$. The states in each $N$-sector can be expressed as 
\begin{equation}\label{JonN}
J_0^{a\{-n_1\}} J_0^{b\{-n_2\}} J_0^{c\{-n_3\}}\ldots|N\rangle,
\end{equation}
where we have denoted the ground state $|0\rangle$ in the sector $N$ as $|N\rangle$. Recall that since we are in the topological limit, the affine generators do not raise the energy level of the ground state, and all the states in the module remain sigma model ground states. 
The number of states in a particular $N$-sector is just given by the number of local observables in that sector, which 
is specified by the dimension of $H_{L^2}^*(\mathcal{M}(\C P^1 \xrightarrow[hol.]{N}\Omega{SU(k)}))$. 

Following \eqref{affinedecomp}, the Hilbert space of a particular $N$-sector can be decomposed as 
\begin{equation}\label{hilbdecomp}
\mathcal{H}^{(top)}_N=\bigoplus_{\widehat{\lambda},\widehat{\mu}}\mathcal{H}^{(top)}_{\widehat{\lambda},\widehat{\mu}},
\end{equation}   
where $\mathcal{H}^{(top)}_{\widehat{\lambda},\widehat{\mu}}$ is a submodule over $\mathfrak{su}(k)_\textrm{aff}$. Since this submodule is finite, we know that it is a subspace of an integrable module. We also know from \eqref{JonN} that this integrable module is a highest weight module. The two previous statements mean that the submodules which form $\mathcal{H}^{(top)}$ are subspaces of dominant highest weight modules \cite{de1997lie}. Following this, we shall take $\widehat{\lambda}$ to be a dominant highest affine weight and $\widehat{\mu}$ to be a dominant affine weight.

A generic state in a dominant highest weight module can be expressed as
\begin{equation}\label{genericstate}
|\widehat{\mu}'\rangle=E^{-\widehat{\alpha}}_{-n}\ldots E^{-\widehat{\beta}}_{-m}|\widehat{\lambda}\rangle.
\end{equation}
Here, $E^{-\widehat{\gamma}}_{-l}$ are lowering operators in the Cartan-Weyl basis of $\mathfrak{su}(k)_{\textrm{aff}}$ that correspond to the complement of its Cartan subalgebra; $|\widehat{\lambda}\rangle$ is a highest weight state associated with a dominant highest affine weight $\widehat{\lambda}=(\overline{\lambda},c_1,i)$; $\widehat{\mu}'=(\overline{\mu},c_1,j)$ is an affine weight in the weight system of the module of dominant highest weight $\widehat{\lambda}$ of level $c_1$  (which is not necessarily dominant); and $\widehat{\alpha}=(\overline{\alpha},c_1,n)$, $\widehat{\beta}=(\overline{\beta},c_1,m)$ are positive affine roots.	Given the state $|\widehat{\lambda}\rangle$, there are several degenerate states $|\widehat{\mu}'\rangle$ which correspond to the weight $\widehat{\mu}'$, each corresponding to a particular choice of positive roots $\widehat{\alpha}\cdots\widehat{\beta}$ which satisfy $\widehat{\mu}'=\widehat{\lambda}-\widehat{\beta}\cdots-\widehat{\alpha}$. 
A Weyl group symmetry maps the affine weight $\widehat{\mu}'$ to the dominant affine weight $\widehat{\mu}$ in the weight system of the same module.

Now, note that the grade of the highest weight in a module is merely a matter of convention. As such, we can shift the grades of $\widehat{\lambda}=(\overline{\lambda},c_1,i)$ and $\widehat{\mu}'=(\overline{\mu},c_1,j)$ to
$\widehat{\lambda}=(\overline{\lambda},c_1,0)$ and $\widehat{\mu}'=(\overline{\mu},c_1,-m)$, where $m=i-j$ is a non-negative integer. In this way, the decomposition \eqref{hilbdecomp} is equivalent to
\begin{equation}
\mathcal{H}^{(top)}_N=\bigoplus_{\widehat{\lambda},\overline{\mu},m}\mathcal{H}^{(top)}_{\widehat{\lambda},\overline{\mu},m},
\end{equation}
with
\begin{equation}
\textrm{dim}\mathcal{H}_{\widehat{\lambda},\overline{\mu},m}^{(top)}
=\textrm{mult}_{\widehat{\lambda}}(\overline{\mu})|_m ,
\end{equation}
where the right hand side indicates the number of degenerate states corresponding to the affine weight $\widehat{\mu}'=(\overline{\mu},c_1,-m)$ in the module of highest weight $\widehat{\lambda}$.
Since $m=i-j$, it follows from \eqref{instadecomp} that the worldsheet instanton number can be written as $N=m+\frac{1}{2}(\overline{\lambda},\overline{\lambda})-\frac{1}{2}(\overline{\mu},\overline{\mu})$.
Note that $m$ is not the eigenvalue of $L_0$, since we only have ground states in a topological theory. One can use the Sugawara construction on the affine Lie algebra generators to find an operator $\widehat{L}_0$ whose eigenvalue is $m$. 

In light of these facts, we may calculate the partition function of the nonpertubative sigma model as
\begin{equation}\label{two}
\begin{aligned}
Z_{(top)}=
&\textrm{Tr}_{\mathcal{H}^{(top)}}p^{\widehat{N}}
\\=&\sum_{N>0}\textrm{dim}\mathcal{H}_N^{(top)}p^N
\\=&\sum_{\widehat{\lambda},\overline{\mu},m}\textrm{dim}\mathcal{H}_{\widehat{\lambda},m,\overline{\mu}}^{(top)}p^{m+\frac{1}{2}(\overline{\lambda},\overline{\lambda})-\frac{1}{2}(\overline{\mu},\overline{\mu})}
\\=&\sum_{\widehat{\lambda}}\sum_{\overline{\mu}}\sum_{m \geq 0} \textrm{mult}_{\widehat{\lambda}}(\overline{\mu})|_m p^{m+h_{\widehat{\lambda}}-\frac{c_{\widehat{\lambda}}}{24}}
\\=&\sum_{\widehat{\lambda}}p^{\frac{\widehat{c}-c_{\widehat{\lambda}}}{24}}\textrm{Tr}_{\widehat{\lambda}}p^{\widehat{L}_0+h_{\widehat{\lambda}}-\frac{\widehat{c}}{24}}
\\=&\sum_{\widehat{\lambda}'}\textrm{Tr}_{\widehat{\lambda}'}p^{\widehat{L}_0+m_{\widehat{\lambda}'}}
\\=&\sum_{\widehat{\lambda}'}\chi^{\widehat{\lambda}'}_{\widehat{su}(k)_{c_1}}(p).
\end{aligned}
\end{equation}

Here, we have the non-negative number
\begin{equation}
h_{\widehat{\lambda}}=\frac{(\overline{\lambda},\overline{\lambda}+2\rho)}{2(c_1+h)},
\end{equation}
and the numbers
\begin{equation}
c_{\widehat{\lambda}}=-12(\overline{\lambda},\overline{\lambda})+12(\overline{\mu},\overline{\mu})+\frac{12(\overline{\lambda},\overline{\lambda}+2\rho)}{c_1+h}
\end{equation}
and
\begin{equation}
\widehat{c}=\frac{c_1 \textrm{dim } \mathfrak{su}(k)}{c_1+h}=\frac{12c_1|\rho|^2}{(c_1+h)h},
\end{equation}
where $\rho$ and $h$ are the Weyl vector and dual Coxeter number of the finite Lie algebra $\mathfrak{su}(k)$. In the 
penultimate line of \eqref{two}, we have shifted the grade of the dominant highest weight $\widehat{\lambda}=(\overline{\lambda},c_1,0)$ to $\widehat{\lambda}'=(\overline{\lambda},c_1,\frac{\widehat{c}-c_{\widehat{\lambda}}}{24})$, whereby
 \begin{equation}
m_{\widehat{\lambda}'}=h_{\widehat{\lambda'}}-\frac{\widehat{c}}{24},
\end{equation} 
and $h_{\widehat{\lambda}'}=h_{\widehat{\lambda}}$, $c_{\widehat{\lambda}'}=c_{\widehat{\lambda}}$. Finally, we obtained a sum over $\chi^{\widehat{\lambda}'}_{\widehat{su}(k)_{c_1}}(p)$, where $\chi^{\widehat{\lambda}'}_{\widehat{su}(k)_{c_1}}(p)$ is the 
character for an irreducible, integrable $\mathfrak{su}(k)_{\textrm{aff}}$ module at level $c_1$ with dominant highest weight $\widehat{\lambda}'$. 

 The partition function for the nonpertubative topological sigma model with $\Omega SU(k)$ target space is thus a sum of characters for modules over $\mathfrak{su}(k)_{\textrm{aff}}$, and therefore transforms like a modular form. From \eqref{bravefinkel}, and since $\textrm{IH}^*(\mathcal{M}_{SU(k)}^N(\R^4)$ for all $N>0$ corresponds to the 1/2 BPS states of the worldvolume theory of a stack of $k$ M5-branes (see Section 5.1), this is also the partition function of the 1/2 BPS sector of the worldvolume theory (using discrete lightcone quantization). 
Note that we have arrived at the same result as ({\cite{Tan2013a}}, equation 3.33) using arguments from quantum field theory (instead of string theory/M-theory).



In the DLCQ gauge, the M5-brane worldvolume contains $S^1\times \R$, where $\R$ is the time coordinate.
Naturally, when one computes the partition function of the M5-brane worldvolume theory (or some BPS sector thereof), which is a trace over identical initial and final states, $S^1\times \R$ becomes equivalent to 
a torus, which we denote $T^2_M$ \cite{Dijkgraaf1998}. In other words, the 1/2 BPS partition function counts the states of the topological sector (i.e., ground states) of the $\mathcal{N}$=(4,4) sigma model on $T^2_M$, with $\mathcal{M}_{SU(k)}^N(\R^4)$ target space, summed over all $N>0$.\footnote{To be precise, computation of the 1/2 BPS partition function in terms of the $\mathcal{N}=(4,4)$ sigma model implies antiperiodic boundary conditions for fermionic fields as we go around the time direction on the torus, $T^2_M$ \cite{hori2003mirror}. Such boundary conditions are not compatible with supersymmetry, since the bosonic fields must be periodic. However, note that since $\mathcal{M}_{SU(k)}^N(\R^4)$ is hyperk\"ahler, and therefore Calabi-Yau, the sigma model with this target space in fact has superconformal symmetry, whereby the supersymmetry transformation parameters are holomorphic or antiholomorphic functions on the worldsheet. One can then choose antiperiodic boundary conditions for these functions, and this choice allows the boundary conditions on the bosonic and fermionic fields to be consistent.} 

These ground states give rise to the spectrum of the 6d $\mathcal{N}$=(2,0) SCFT \cite{Dijkgraaf1996,dijkgraaf1997bps}, now effectively on $T^2_M \times \R^4$, where this is the same as 4d $\mathcal{N}$=4 Super Yang-Mills theory (SYM) on $\R^4$ \cite{witten1995some,Witten2007}. 
The Montonen-Olive duality of 4d $\mathcal{N}$=4 SYM can then be understood as modular covariance of the affine Lie algebra characters in \eqref{two}. 

It is also known that the Hilbert space of the 1/2 BPS sector of a single M5-brane is the Fock space of a 2d free chiral scalar CFT \cite{Dijkgraaf1998,Dijkgraaf1998long},\footnote{Note that a chiral scalar field, $\varphi$, has a self-dual `field strength', $d\varphi$, just like the field strength of the 2-form potential in the 6d $\mathcal{N}=(2,0)$ SCFT \cite{Witten1997}. } wherein the Laurent nonzero modes of the scalar field generate the Heisenberg algebra, and this agrees with our 1/2 BPS partition function \eqref{two} for $G=U(1)$, which is the sum of characters for Heisenberg algebra modules. 

Since the 1/2 BPS sector corresponds to the topological sector of the $\mathcal{N}=(4,4)$ sigma model with $\mathcal{M}_{SU(k)}^N(\R^4)$ target space, which in turn is given by the $Q$-cohomology of ground states of supersymmetric quantum mechanics on $\mathcal{M}_{SU(k)}^N(\R^4)$, we do not see any `stringy' effects of the little strings in the 1/2 BPS partition function. This situation will change, as we shall see, in the 1/4 BPS case.


\subsection{1/4 BPS Sector}

For the quasi-topological sigma model, we calculate a generalization of the partition function given in $\eqref{topppf}$, i.e.,
\begin{equation}\label{threeprelude}
Z_{(q.t.)}=\Tr_{\mathcal{H}^{(q.t.)}}(p^{\widehat{N}}\otimes q^{\widehat{\widehat{L}}_0}),
\end{equation} 
where $\widehat{N}$ is the instanton number operator previously defined,   $\widehat{\widehat{L}}_0$ is a grading operator one can find via a generalization of the Sugawara construction to the case of toroidal Lie algebras, and $q=e^{2\pi i \tau}$ is a complex parameter.
 $\widehat{\widehat{L}}_0$ grades the elements of the toroidal Lie algebra via
\begin{equation}\label{torsuga}
[\widehat{\widehat{L}}_0,J^{ak}_m]=-mJ^{ak}_{m}.
\end{equation}
The quasi-topological partition function \eqref{threeprelude} is a natural generalization of the topological partition function $\eqref{topppf}$, since it reduces precisely to the latter when we take the topological limit, whereby the toroidal Lie algebra reduces to an affine Lie algebra.  

Recall from \eqref{module2} and below that the Hilbert space $\mathcal{H}^{(q.t.)}$ of the quasi-topological model is made up of submodules (labelled by $N$) over the toroidal Lie algebra $\mathfrak{su}(k)_{\textrm{tor}}$, i.e., they are of the form  
\begin{equation}
J_{-m_1}^{a\{-n_1\}} J_{-m_2}^{b\{-n_2\}} J_{-m_3}^{c\{-n_3\}}\ldots|N\rangle.
\end{equation}
Also recall that, unlike the fully topological case, the toroidal Lie algebra generators actually raise the energy level of a state, for $m_i \neq 0$. The states in each $N$-sector of $\mathcal{H}^{(q.t.)}$ can be written as 
\begin{equation}\label{quastopstates}
\begin{aligned}
J_{-m_1}^{a\{-n_1\}} J_{-m_2}^{b\{-n_2\}} J_{-m_3}^{c\{-n_3\}}\ldots|N\rangle &=|N,m_1+m_2+m_3\ldots=l\rangle
\\&=|N,l\rangle,
\end{aligned}
\end{equation}
where $l$ denotes the eigenvalue of $L_0$. From \eqref{energyraising2} and \eqref{torsuga}, we see that the eigenvalue of $L_0$ is the same as the eigenvalue of $\widehat{\widehat{L}}_0$.\footnote{This does not mean that $L_0$ and $\widehat{\widehat{L}}_0$ are identical, since they are the Laurent zero-modes of different spin-2 fields.} As before, we are only interested in $N>0$.

Expressing the states of $\mathcal{H}^{(q.t.)}$ as in \eqref{quastopstates}, we have
\begin{equation}\label{three}
\begin{aligned}
Z_{(q.t.)}=
&\Tr_{\mathcal{H}^{(q.t.)}}(p^{\widehat{N}}\otimes q^{\widehat{\widehat{L}}_0})
\\=&\sum_{N>0,l=0} \langle N,l|p^{\widehat{N}}\otimes q^{\widehat{\widehat{L}}_0}|N,l\rangle
\\=&\sum_{N>0} \langle N|p^{\widehat{N}}|N\rangle \sum_{l=0}\langle l|q^{\widehat{\widehat{L}}_0}| l\rangle
\\=&\sum_{N>0} \textrm{dim}\mathcal{H}_N^{(top)} p^{N} \sum_{l=0} P(l) q^{l} 
\\=&\sum_{\widehat{\lambda}}\chi^{\widehat{\lambda}}_{\widehat{su}(k)_{c_1}}(p)\prod_{l=1}\frac{1}{1-q^l}
\\=&q^{\frac{1}{24}}\sum_{\widehat{\lambda}}\chi^{\widehat{\lambda}}_{\widehat{su}(k)_{c_1}}(p)\frac{1}{\eta(\tau)}.
\end{aligned}
\end{equation}
Here, $P(l)$ is the number of partitions of the integer $l$. In the third equality, we have split the states as tensor products since $\widehat{N}$ and $\widehat{\widehat{L}}_0$ 
act independently of one another.
We have also made use of our 1/2 BPS result (equation $\eqref{two}$). Hence, we obtain a sum of characters for $\mathfrak{su}(k)_{\textrm{aff}}$ modules multiplied by a Virasoro character. 

This is the partition function for the nonpertubative quasi-topological sigma model with $\Omega SU(k)$ target space. 
Since the states of this nonpertubative quasi-topological theory correspond to the 1/4 BPS states of the worldvolume theory of a stack of $k$ M5-branes (see Section 5.1 and the penultimate paragraph of Section 3.1), $\eqref{three}$ is also the partition function for the 1/4 BPS sector of the worldvolume theory (using discrete lightcone quantization). Each factor in the partition function transforms like a modular form, and therefore, the partition function transforms like an automorphic form for $SO(2,2;\Z)$ since
\begin{equation}
SO(2,2;\Z)\cong SL(2,\Z) \times SL(2,\Z).
\end{equation}
The 1/4 BPS quantum worldvolume theory is equivalent, via inspection of \eqref{three}, to  chiral WZW model $\times$ chiral free boson $\times$ interactions. 

By analyzing the 1/4 BPS sector of the M5-brane worldvolume theory via our quasi-topological sigma model, we have gone beyond the 6d $\mathcal{N}$=(2,0) SCFT, enabling us to see the `stringy' effects of the little strings, i.e., the partition function is enhanced from one which transforms like a modular form in the 1/2 BPS case to one which transforms like an automorphic form for $SO(2,2,\mathbb{Z})$, which is just the T-duality group for the worldvolume torus, $T^2_M$, that appears when taking the trace in the M5-brane partition function.\footnote{The little strings propagating in a torus can have nontrivial Kaluza-Klein momentum as well as nontrivial winding along each dimension of the torus. In other words, the string states can have two winding number charges and two Kaluza-Klein charges. The $SO(2,2;\Z)$ symmetry arises as the rotational symmetry of the even, self-dual Narain lattice $\Gamma^{2,2}$ of signature (2,2) formed by these charges.}

It is also worth noting that based on our analysis, the 1/4 BPS partition function \eqref{three} of the M5-brane worldvolume theory basically counts the dimensions of \v{C}ech cohomology classes on the sheaf of chiral de Rham complex on $\mathcal{M}_{SU(k)}^N(\R^4)$, and we see that this transforms like an automorphic form for $SO(2,2;\Z)$. In other words, our results lead us to a relationship between a sheaf of supersymmetric vertex algebras and an object that transforms like an automorphic form.

\acknowledgments
We would like to thank Michael Atiyah, Kimyeong Lee and Graeme Segal for various helpful discussions. This work is supported by NUS Tier 1 FRC Grant R-144-000-316-112.


  
\end{document}